\documentclass[journal]{IEEEtran}

\ifCLASSINFOpdf
   \usepackage[pdftex]{graphicx}
    \usepackage{epstopdf}
\else
\fi

\usepackage{amssymb}
\usepackage{amsmath}
\usepackage{amsthm}
\newtheorem{theorem}{Theorem}

\usepackage{xcolor}
\usepackage{cite}

\begin{document}
%
\title{An auto-parameter denoising method for nuclear magnetic resonance spectroscopy based on low-rank Hankel matrix}

\author{Tianyu~Qiu, 
        Wenjing~Liao, 
        Di~Guo,
        Dongbao~Liu,
        Xin~Wang,
        Jian-Feng~Cai,
        and~Xiaobo~Qu$ ^{*} $
\thanks{This work was supported in part by National Natural Science Foundation of China (61971361, 61871341, 61811530021, 61672335),  National Key R{\&}D Program of China
(2017YFC0108703), Natural Science Foundation of Fujian Province of China (2018J06018), Fundamental Research Funds for the Central Universities (20720180056), Science and
Technology Program of Xiamen (3502Z20183053), and China Scholarship Council. Asterisk indicates corresponding author (Email: quxiaobo@xmu.edu.cn)}
\thanks{Tianyu Qiu, Dongbao Liu, Xin Wang and Xiaobo Qu are with the Department of Electronic Science, Fujian Provincial Key Laboratory of Plasma and Magnetic Resonance, Xiamen University, Xiamen, China.}
\thanks{Wenjing Liao is with School of Mathematics, Georgia Institute of Technology, Atlanta, GA, USA.}
\thanks{Di Guo is with School of Computer and Information Engineering, Fujian Provincial University Key Laboratory of Internet of Things Application Technology, Xiamen University of Technology, Xiamen, China.}
\thanks{Jian-Feng Cai is with Department of Mathematics, Hong Kong University of Science and Technology, Hong Kong, China.}}

%
%

\markboth{IEEE TRANSACTIONS ON BIOMEDICAL ENGINEERING,~Vol.~, No.~, ~}%
{Shell \MakeLowercase{\textit{et al.}}: Bare Demo of IEEEtran.cls for IEEE Journals}
%



\maketitle

\begin{abstract}
Nuclear Magnetic Resonance (NMR) spectroscopy, which is modeled as the sum of damped exponential signals, has become an indispensable tool in various scenarios, such as the structure and function determination, chemical analysis, and disease diagnosis. NMR spectroscopy signals, however, are usually corrupted by Gaussian noise in practice, raising difficulties in sequential analysis and quantification of the signals. The low-rank Hankel property plays an important role in the denoising issue, but selecting an appropriate parameter still remains a problem. In this work, we explore the effect of the regularization parameter of a convex optimization denoising method based on low-rank Hankel matrices for exponential signals corrupted by Gaussian noise. An accurate estimate on the spectral norm of weighted Hankel matrices is provided as a guidance to set the regularization parameter. The bound can be efficiently calculated since it only depends on the standard deviation of the noise and a constant. Aided by the bound, one can easily obtain an auto-setting regularization parameter to produce promising denoised results. Our experiments on synthetic and realistic NMR spectroscopy data demonstrate a superior denoising performance of our proposed approach in comparison with the typical Cadzow and the state-of-the-art QR decomposition methods, especially in the low signal-to-noise ratio regime.
\end{abstract}


%
\IEEEpeerreviewmaketitle

\section{Introduction}
%
%
%
%
\IEEEPARstart{N}{uclear} magnetic resonance (NMR) spectroscopy has grown into an essential tool for biomedical studies, such as the structure determination \cite{2009_Nature_Inomata}, metabolic analysis \cite{ 2010_NatureP_Beckonert}, and medical diagnosis \cite{1996_NatureM_Preul}. However, NMR spectroscopy signals are often corrupted by noise during acquisition and/or transmission. The noise problem turns out to be severe in the low Signal-to-Noise Ratio (SNR) regime \cite{2014_SSNMR_Man, 2019_TMI_Fan}. Therefore, there is a strong demand to denoise signals, particularly in the low SNR regime. 

Gaussian noise is commonly encountered in NMR spectroscopy denoising applications \cite{2014_PNAS_Delsuc, 2014_MRM_ZPLiang, 2018_TSP_Ying, 1997_TBME_Tufts}. One of the most effective and widely adopted approaches to suppress Gaussian noise is to average multiple signal acquisitions. However, the multiple acquisitions are not always available or too costly in real applications. For this reason, effective denoising of the signals with a limited number of scans is favorable.

Numerous efforts have been made to denoise NMR spectroscopy signals. Among them, exploiting the exponential characteristic of NMR spectroscopy signals has been grown into a powerful tool \cite{1988_TASSP_Cadzow, 1993_JMR_LPHwang, 2010_Gillard_CadzowAlgorithm}. Such low-rank properties were also utilized in the field of NMR spectroscopy reconstruction \cite{2015_Angew_Xiaobo,2018_TSP_Ying,2019_Anie_Qu} and magnetic resonance spectroscopic imaging \cite{2013_TBME_ZPLiang,2019_MRI_Noiseworthy,2016_MRM_Cao}. The Cadzow enhancement approach is popular in spectra denoising with the exploitation of the low-rank property of exponentials \cite{1988_TASSP_Cadzow, 1993_JMR_LPHwang, 2010_Gillard_CadzowAlgorithm}. Compared with some typical denoising methods, such as the smoothing approach \cite{1980_PNMR_Lindon}, wavelet thresholding \cite{1995_TIT_Donoho, 1997_JMR_Barache}, Maximum entropy \cite{1990_PNAS_Hoch}, and covariance matrix \cite{2015_ARNMRS_Takeda, 2007_BBAB_Glaubitz}, Cadzow method is more theoretically adopted to the denoising of all NMR spectroscopy signals. However, it is a challenging task to choose a proper number $R$ of exponential components in practical applications, unless a priori information is given. Efforts have been made to estimate $R$, such as the indicator function \cite{1977_AC_Malinowski} and the significance level function \cite{1999_JCACS_Malinowski}, but the estimation of $R$ may not be satisfactory enough to yield good results \cite{2019_ASR_Laurent}. Another denoising method called random QR denoising method (rQRd) is based on an approximate low-rank decomposition, and accelerates the computation by avoiding the Singular Value Decomposition (SVD) in the Cadzow method \cite{2014_PNAS_Delsuc}. It is, however, also based on an  estimation of the rank $R$. 

\begin{figure}[ht]
  \centering
  \includegraphics[width=3.0in]{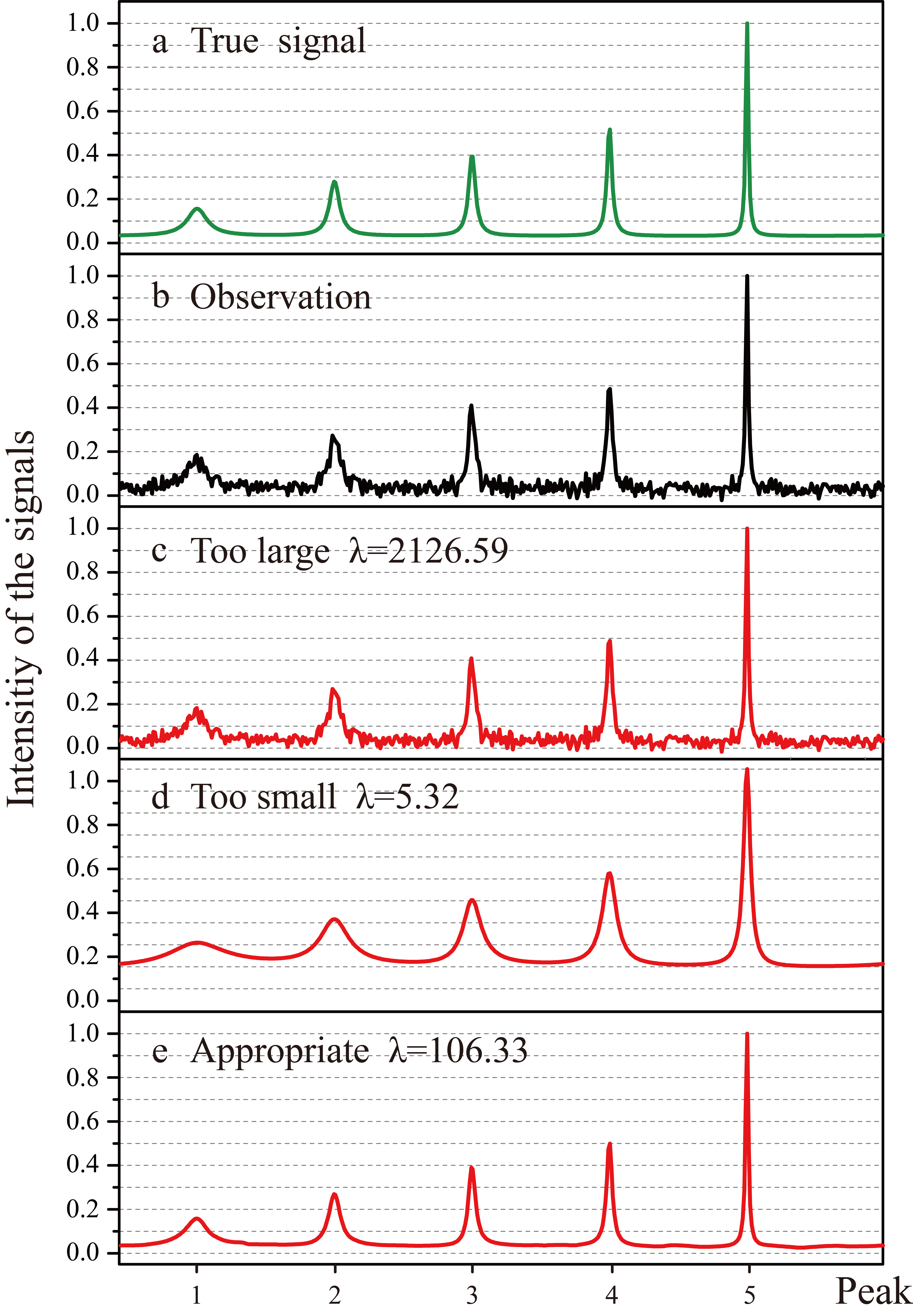}
  \caption{A denoising example with different choices of $\lambda$. 
   (a) The true signal. (b) The observation with Gaussian noise ($\sigma=0.02$). (c)-(e) The denoised results with $\lambda = 2126.59$, $5.32$, and $106.33$, respectively. Note: without explicit illustration, $\lambda$ is in the data consistency term in this paper.
  }\label{fig:1}
\end{figure}

This low-rank Hankel property also can be exploited in an unconstrained convex optimization method for the reconstruction issue \cite{2014_TIT_Chi, 2015_Angew_Xiaobo}. The method, which is named as Low-Rank Hankel Matrix reconstruction method (LRHM), also can be used for denoising, and one may receive a good result. The regularization parameter $\lambda$ plays an important role in the results. As an example, Fig. \ref{fig:1} shows the denoising results with different $\lambda$. If $\lambda$ is too large, the majority of the noise remains since the effect of the nuclear norm minimization is ignorable; if $\lambda$ is too small, the spectral peaks are seriously distorted. Unfortunately, the choice of $\lambda$ is still based on users' experience. Exploring the effect and the proper choice of $\lambda$ is still of great demand and challenging.

In this paper, we explore the effect of the regularization parameter, and show that a good $\lambda$ can be automatically chosen according to the spectral norm of a weighted Hankel matrix, which is estimated by random matrix theory as a guideline for the selection of a proper $\lambda$. One only needs to estimate the standard deviation of the noise, which also can be set automatically, to calculate this proper $\lambda$. Numerical experiments on both synthetic and real NMR spectroscopy data show that noise can be effectively removed when the parameter is chosen according to our analysis.

The rest of the paper is organized as follows. Section {\ref{section:2}} briefly reviews the signal model of NMR spectroscopy signals and LRHM in the denoising issue. Section {\ref{section:3}} is deveoted to analyzing the selection of $\lambda$ and estimating the spectral norm of weighted Hankel matrices. Section {\ref{section:4}} contains numerical results on synthetic and real NMR spectroscopy data. Section {\ref{section:5}} discusses the robustness to the estimate on the noise standard deviation. Finally,  we conclude and discuss future works in Section {\ref{section:6}}.

Notations used in the paper are introduced below. We denote vectors through bold lowercase letters and matrices through bold uppercase letters. The entry in vectors and matrices is denoted by a normal letter with a subscript which stands for its location. For example, $x_n$ denotes the $n^{th}$ element of $\mathbf{x}$, and $X_{m,n}$ denotes the $(m,n)^{th}$ entry of $\mathbf{X}$. For any vector $\mathbf{x}$, $\left\|{\mathbf{x}}\right\|_2$ represents the $l_2$ norm. For any matrix $\mathbf{X}$, $\left\|{\mathbf{X}}\right\|_*$ and $\left\|{\mathbf{X}}\right\|_2$ denote the nuclear norm and the spectral norm, respectively. The Hadamard product is denoted by $\circ$. We use superscript $T$ and $H$ to denote the transpose and the conjugate transpose of $\mathbf{x}$ and $\mathbf{X}$.
Most of operators are denoted by calligraphic letters. We denote $diag$ as the operator transforming a sequence to a diagonal matrix whose diagonal entries are given by the sequence.

\section{Connection to prior work}\label{section:2}

In the time domain, NMR spectroscopy signal, which is named as Free Induction Decay (FID), can be expressed as the sum of $ R $ exponentials:
\begin{equation}\label{eq:1}
  x_{0}(t_{n})=\sum_{r=1}^{R}a_{r}e^{\left({j2{\pi}f_{r}-{\tau}_{r}}\right)t_{n}}, \ n=0,\ldots,2N
\end{equation}
where $ a_{r} $ denotes the signal amplitude, $ f_{r} $ is the central frequency, and $ {\tau}_{r} $ is the decay factor. The number of exponentials $R$ is usually small.

In practice, observations are often contaminated by noise and one receives $\mathbf{y}=\mathbf{x}_{0}+\mathbf{z}$, where $\mathbf{x}_0 = \{x_0(t_n)\}_{n=0}^{2N}$ is a noiseless signal and $\mathbf{z}\in\mathbb{C}^{2N+1}$ is a random vector whose real and imaginary parts are i.i.d Gaussian with mean 0 and variance ${\sigma}^2$.

Exponential signals can be transformed into Hankel matrices with a Vandermonde decomposition. Given $\mathbf{x}_{0}$, one forms the square Hankel matrix
$$
\mathcal{R}\mathbf{x}_{0} =
 \left[
\begin{array}{cccc}
  x_{0}\left({t_0}\right) & x_{0}\left({t_1}\right) & \cdots & x_{0}\left({t_{N}}\right) \\
  x_{0}\left({t_1}\right) & x_{0}\left({t_2}\right) & \cdots & x_{0}\left({t_{N+1}}\right) \\
  \vdots & \vdots & \vdots & \vdots \\
  x_{0}\left({t_{N}}\right) & x_{0}\left({t_{N+1}}\right) & \cdots & x_{0}\left({t_{2N}}\right)
\end{array}
\right],
$$ where $\mathcal{R}: \mathbb{C}^{2N+1} \rightarrow \mathbb{C}^{(N+1) \times (N+1)} $ is the operator transforming a vector to the square Hankel matrix. It is well known that $\mathcal{R}\mathbf{x}_{0}$ is of rank $R$ \cite{1996_Hoch_NMR, 1999_PNMRS_Koehl}. 

The denoising method we explore is based on the low-rank property of $\mathcal{R}\mathbf{x}_{0}$\cite{2015_Angew_Xiaobo}, and called Convex Hankel lOw-Rank matrix approximation for Denoising exponential signals (CHORD), where one solves the following optimization problem:
\begin{equation}\label{eq:2}
  \mathbf{\hat{x}}=\arg\min_{\mathbf{x} \in \mathbb{C}^{2N+1}} \left\|{\mathcal{R}\mathbf{x}}\right\|_{*}+\frac{\lambda}{2}\left\|{\mathbf{y}-\mathbf{x}}\right\|_{2}^{2},
\end{equation}
where $\lambda$ denotes the regularization parameter, $\hat{\mathbf{x}}$ denotes the minimizer. The nuclear norm $\left\|{\cdot}\right\|_{*}$ is a surrogate for the rank \cite{2010_SIAM_JFCai}.

Alternating Direction Method of Multipliers (ADMM) \cite{2011_FTM_Boyd} is a typical iterative algorithm, which can be used to solve (\ref{eq:2}). 

The optimization problem in (\ref{eq:2}) involves a single regularization parameter $\lambda$, and the denoised result crucially depends on the choice of $\lambda$. Therefore, setting an appropriate $ \lambda $ is a crucial issue in this denoising method. This paper provides an automatic estimate on the proper choice of $\lambda$, and validations by experimental results.

\section{An automatic estimate of the regularization parameter $\lambda$}\label{section:3}

This section provides an estimate of the proper $\lambda$ through establishing a relation between $\lambda$ and the spectral norm of weighted Hankel matrices.

As $\mathbf{\hat{x}}$ is the minimizer of (\ref{eq:2}), the subgradient of (\ref{eq:2}) vanishes at $\hat{\mathbf{x}}$. According to the subgradient of the nuclear norm \cite{1992_Watson_LAA, 2003_Bertsekas_Conv, 2010_SIAM_JFCai, 2010_ProIEEE_Candes}, the subgradient of (\ref{eq:2}) is derived as

\begin{equation}\label{eq:4}
  \lambda\left({\mathbf{x}_{0}+\mathbf{z}-\mathbf{\hat{x}}}\right)=\mathcal{R}^{*}\left({\mathbf{\hat{U}}\mathbf{\hat{V}}^{H}+\mathbf{\hat{W}}}\right),
\end{equation}
where the matrices $\mathbf{\hat{U}}$, $\mathbf{\hat{V}}{\in}\mathbb{C}^{(N+1)\times(N+1)}$ are from the SVD of $\mathcal{R}\mathbf{\hat{x}}$ such that $\mathcal{R}\mathbf{\hat{x}}=\mathbf{\hat{U}}\mathbf{\hat{\Sigma}}\mathbf{\hat{V}}^{H}$, and $\mathbf{\hat{W}}{\in}\mathbb{C}^{(N+1)\times(N+1)}$ satisfies $ \mathbf{\hat{U}}^{H}\mathbf{\hat{W}}=\mathbf{0} $, $ \mathbf{\hat{W}}\mathbf{\hat{V}}=\mathbf{0} $, and $ \left\|\mathbf{\hat{W}}\right\|_2\leq 1 $. $\mathcal{R}^{*} : \mathbb{C}^{(N+1)\times(N+1)} \rightarrow \mathbb{C}^{2N+1}$ is an operator transforming a matrix into vector via summing each anti-diagonal.

Denote the vector $\mathbf{w}$ is the weights defined as $ \mathbf{w}=\left[{
\begin{array}{ccccccc}
  1 & 2 & \cdots & N+1 & \cdots & 2 & 1
\end{array}
}\right]^{T}\in\mathbb{R}^{2N+1}$ and the symbol $\circ$ stands for Hadamard product.

Since $\left({\lambda\mathcal{R}\frac{1}{\mathbf{w}}\circ\left({\mathbf{x}_{0}+\mathbf{z}-\mathbf{\hat{x}}}\right)}\right)$ is an approximation of $\mathbf{\hat{U}}\mathbf{\hat{V}}^{H}+\mathbf{\hat{W}}$, the proper $\lambda$ is chosen as below

\begin{equation}\label{eq:7}
  \frac{1}{\left(\left|\left\|\mathbf{Z}\right\|_{2}+\left\|\mathbf{\tilde{X}}\right\|_{2}\right|\right)} \leq \lambda \leq {\frac{1}{\left(\left|\left\|\mathbf{Z}\right\|_{2}-\left\|\mathbf{\tilde{X}}\right\|_{2}\right|\right)}},
\end{equation}
where $\mathbf{Z}=\left(\mathcal{R}\frac{1}{\mathbf{w}}\right)\circ\mathcal{R}\mathbf{z}$ denotes a weighted Hankel matrix such that

\begin{equation}\label{eq:6}
  \mathbf{Z}=\left(\mathcal{R}\frac{1}{\mathbf{w}}\right)\circ\mathcal{R}\mathbf{z}=
  \left(
  \begin{array}{cccc}
    z_1 & \frac{z_2}{2} & \cdots & \frac{z_{N+1}}{N+1} \\
    \frac{z_2}{2} & \frac{z_3}{3} & \cdots & \frac{z_{N+2}}{N} \\
    \vdots & \vdots & \cdots & \vdots \\
    \frac{z_{N+1}}{N+1} & \frac{z_{N+1}}{N+1} & \cdots & z_{2N+1}
  \end{array}
  \right),
\end{equation}
and $\mathbf{\tilde{X}}$ denote $\mathbf{\tilde{X}}=\left(\mathcal{R}\frac{1}{\mathbf{w}}\right)\circ\mathcal{R}\left(\mathbf{x}_{0}-\mathbf{\hat{x}}\right)$.

\begin{figure}[ht]
  \centering
  \includegraphics[width=2.5in]{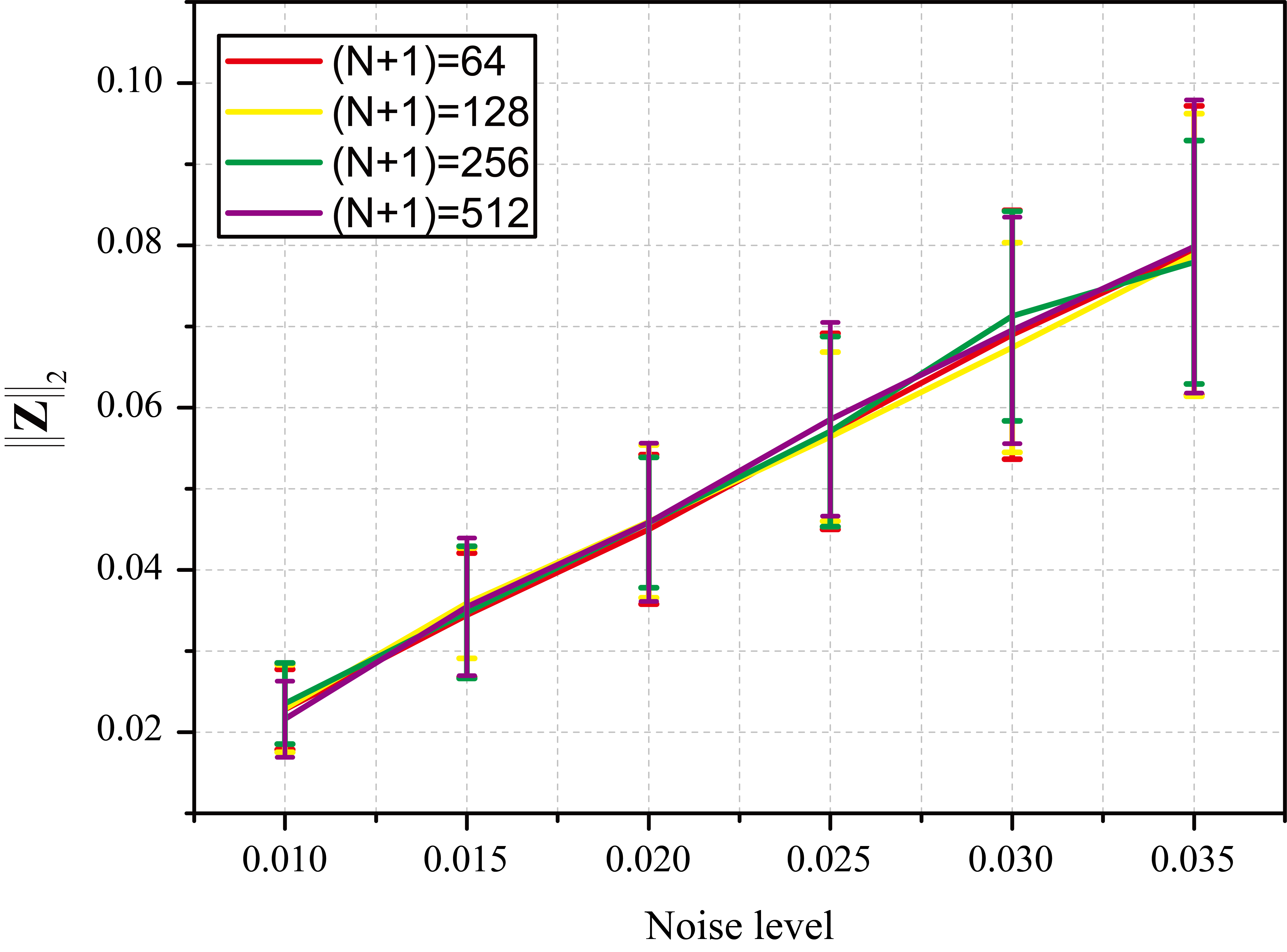}
  \caption{
  The relation between $\left\|\mathbf{Z}\right\|_{2}$ and the standard deviation $\sigma$ of the Gaussian noise $ \mathbf{z} $ in 100 Monte Carlo trials. The Matrix $\mathbf{Z}$ is of size $ (N+1){\times}(N+1) $ with $(N+1)=64$, $128$, $256$, $512$, respectively. The curve represents the mean of $\left\|\mathbf{Z}\right\|_{2}$ in 100 trails versus $\sigma$, and the standard deviation of $\left\|\mathbf{Z}\right\|_{2}$ in $100$ trails is indicated by the vertical bar. 
  }\label{fig:2}
\end{figure}

\begin{figure}[ht]
  \centering
  \includegraphics[width=2.5in]{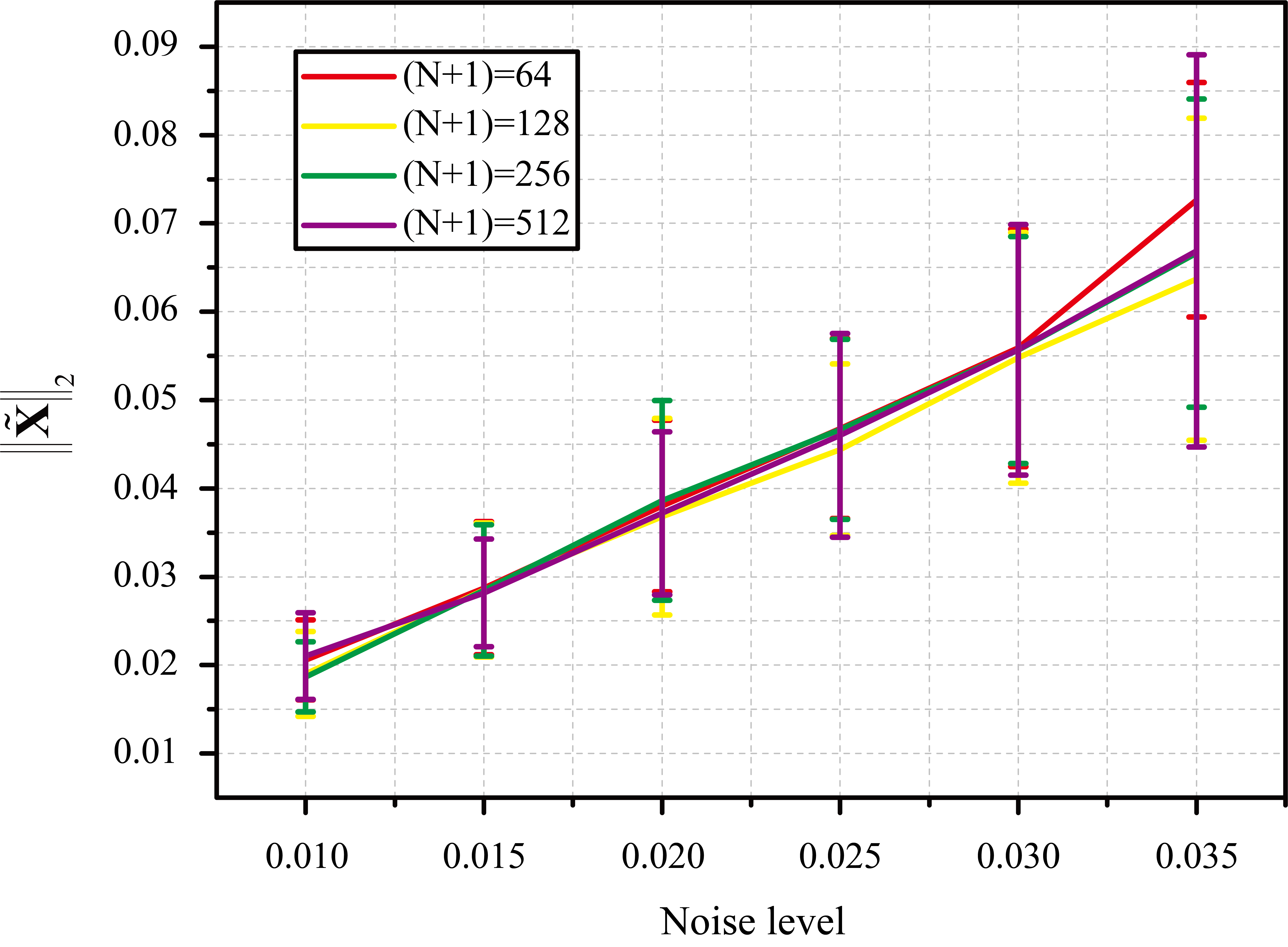}
  \caption{
  The relation between $\left\|\mathbf{\tilde{X}}\right\|_{2}$ and the standard deviation $\sigma$ of the Gaussian noise $ \mathbf{z} $ in 50 Monte Carlo trials. The Matrix $\mathbf{\tilde{X}}$ is of size $ (N+1){\times}(N+1) $ with $(N+1)=64$, $128$, $256$, $512$, respectively. The curve represents the mean of $\left\|\mathbf{\tilde{X}}\right\|_{2}$ in $50$ trails versus $\sigma$, and the standard deviation of $\left\|\mathbf{\tilde{X}}\right\|_{2}$ in 50 trails is indicated by the vertical bar. Note: $\mathbf{x}_0$ are damped exponential signals with random $a_r$, $f_r$ and ${\tau}_r$. $\mathbf{\hat{x}}$ is obtained from CHORD.
  }\label{fig:3}
\end{figure}

In order to explore the relationship among the spectral norm of weighted Hankel matrices, the noise level and the size of matrix, we did sufficient Monte Carlo trials on synthetic data and Gaussian noise. Results in Fig. \ref{fig:2} and Fig. \ref{fig:3} show that the empirical means of $\left\|\mathbf{Z}\right\|_2$ and $\left\|\mathbf{\tilde{X}}\right\|_2$ are almost independent of $N$. Furthermore, these empirical means increase as the increasing of the standard deviation $\sigma$ of the noise. 

In applications, we hope to preserve signal details as much as possible, we propose to select the regularization parameter as

\begin{equation}\label{eq:8}
{\lambda}^{*}=\frac {1} {\left|\mathbb{E}\left\|\mathbf{Z}\right\|_2-\mathbb{E}\left\|\mathbf{\tilde{X}}\right\|_2\right|},
\end{equation}
where the symbol $\mathbb{E}$ denotes the expectation.

In order to provide a proper choice of $\lambda$, we estimate an upper and lower bound of $\mathbb{E}\left\|\mathbf{Z}\right\|_2$. With respect to $\mathbb{E}\left\|\mathbf{\tilde{X}}\right\|_2$, we provide an empirical value based on sufficient numerical experiments on synthetic data.

\subsection{The bounds of $\mathbb{E}{\left\|{\mathbf{Z}}\right\|_{2}}$}

Actually, for estimating bounds of the spectral norm of Hankel matrices given by random vectors, numerical achievements have been made \cite{2007_ECP_Meckes, 2013_Math_Meckes, 2015_FnTML_Tropp, 2015_TSP_Wenjing}. In this subsection, we focus on estimating bounds of the spectral norm of weighted Hankel matrices. Theorem \ref{theorem:1} and \ref{theorem:2} provide a lower and upper bounds of $\mathbb{E}\left\|\mathbf{Z}\right\|_2$, respectively. All details of proofs and the asymptotic analysis have been presented in Supplementary.

\begin{theorem}\label{theorem:1}
Suppose the real and imaginary parts of the entries in $ \mathbf{z}{\in}\mathbb{C}^{2N+1} $ are i.i.d. Gaussian random variables with mean 0 and variance $ {\sigma}^{2} $. Define $R_{N}$ and $Q_{N}$ such that
\begin{equation}\label{eq:9}
    R_{N}^{2}=\sum_{k=0}^{2N}\left|{d_k}\right|^{2} \text{and } Q_{N}^{4}=\sum_{k=0}^{2N}\left|{d_k}\right|^{4},
\end{equation}
where $ d_{k}=
  \left\{
  \begin{array}{ll}
    \frac{2}{(k+1)(k+2)}\sum_{m=0}^{k}\frac{1}{m+1}, 0\leq k\leq N \\[1mm]
    \frac{2}{(2N-k+1)(k+2)}\sum_{m=k}^{2N}\frac{1}{m-N+1}, N<k\leq 2N \\[1mm]
  \end{array}
  \right. $. \\[1mm]
Then there exists a constant $C$ such that the matrix $ \mathbf{Z}$ defined in (\ref{eq:7}) satisfies
  \begin{equation}\label{eq:10}
    \mathbb{E}{\left\|{\mathbf{Z}}\right\|_{2}}
  \geq\sigma\frac{C(N+1)}{2N+1}\sqrt{R_{N}^{2}\left({1+\log\frac{R_{N}^{4}}{Q_{N}^{4}}}\right)}.
  \end{equation}
\end{theorem}

\begin{theorem}\label{theorem:2}
  Suppose the real and imaginary parts of the entries in $ \mathbf{z}{\in}\mathbb{C}^{2N+1} $ are i.i.d. Gaussian random variables with mean 0 and variance $ {\sigma}^{2} $. Then 
  \begin{equation}\label{eq:11}
    \mathbb{E}\left\|{\mathbf{Z}}\right\|_{2}{\leq}{\sigma}\sqrt{2C_{\mathbf{w}}\log\left({2N+2}\right)},
\end{equation}
where $C_{\mathbf{w}} = \max(\sum_{k=0}^{N}{w_{k}^{-2}},\sum_{k=1}^{N+1}{w_{k}^{-2}},\ldots,\sum_{k=N}^{2N}{w_{k}^{-2}})$ with the vector $\mathbf{w}$ defined in (\ref{eq:6}).

\end{theorem}

Two theorems above provide the following upper and lower bounds of $\mathbb{E}\left\|{\mathbf{Z}}\right\|_{2}$:

\begin{footnotesize}
\begin{equation}\label{eq:12}
  {\sigma}\frac{C(N+1)}{2N+1}\sqrt{R_{N}^{2}\left({1+\log\frac{R_{N}^{4}}{Q_{N}^{4}}}\right)}{\leq}\mathbb{E}\left\|{\mathbf{Z}}\right\|_{2}{\leq}{\sigma}\sqrt{2C_{\mathbf{w}}\log\left({2N+2}\right)}.
\end{equation}
\end{footnotesize}

The upper bound scales as $\sigma\sqrt{\log N}$, while the lower bound depends on $R_{N}$ and $Q_{N}$. When $N$ is large enough, the upper bound and the lower bound only differ by a factor of $\sqrt{\log N}$. Therefore, we suggest to choose $\mathbb{E}\left\|\mathbf{Z}\right\|_{2}$ as

\begin{equation}\label{eq:13}
  \mathbb{E}\left\|\mathbf{Z}\right\|_{2}=\frac{C(N+1)}{(2N+1)}\sqrt{R_{N}^{2}\left({1+\log\frac{R_{N}^{4}}{Q_{N}^{4}}}\right)}{\sigma}.
\end{equation}

We next find the empirical constant $C$ through repetitive experiments on synthetic data. According to Theorem \ref{theorem:1}, $C>0$ is a constant, which is independent of the signal length and the standard deviation $\sigma$. 

\begin{figure}[ht]
  \centering
  \includegraphics[width=3.5in]{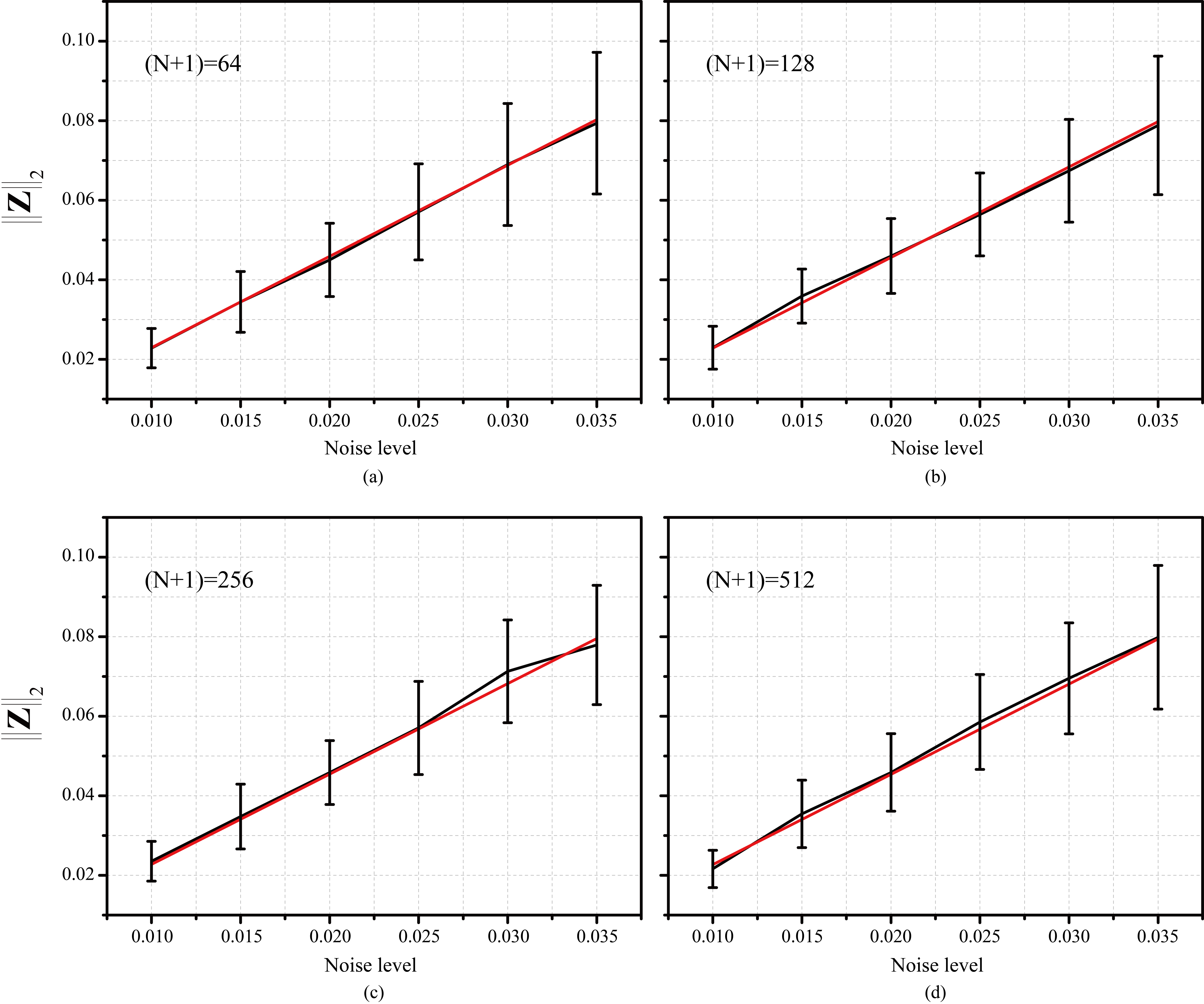}
  \caption{
  $\mathbb{E}\left\|\mathbf{Z}\right\|_2$ and the lower bound with the suggested C under different matrix sizes. The Matrix $\mathbf{Z}$ is of size $ (N+1){\times}(N+1) $ with $(N+1)=64$, $128$, $256$, and $512$, respectively. The vertical axis denotes the value of $\left\|\mathbf{Z}\right\|_2$ and the horizontal axis denotes the standard deviation of Gaussian noise. The black curves stand for the empirical mean of $\left\|\mathbf{Z}\right\|_2$ in Fig. \ref{fig:2}. Red lines denote the lower bounds with $C=2.9$.
  }\label{fig:4}
\end{figure}

We use a series of C to approximate the results in Fig. \ref{fig:2}, and suggest $ C = 2.9 $ for denoising. Results in Fig. \ref{fig:4} confirms that the conclusion in Theorem \ref{theorem:1} is well capable of estimating $\mathbb{E}\left\|\mathbf{Z}\right\|_2$.

\subsection{The empirical $\mathbb{E}\left\|\mathbf{\tilde{X}}\right\|_2$}\label{{subsection3.6}}

This subsection is devoted to an empirical estimate of $\mathbb{E}\left\|\mathbf{\tilde{X}}\right\|_2$. We perform experiments with different $N$, $\sigma$, $\lambda$, signals and noises in order to determine a proper empirical estimate value.

To evaluate the denoising performance, we define the following Relative Least Normalized Error (RLNE) as the objective criteria
\begin{equation}\label{eq:14}
  RLNE=\frac{\left\|{\mathbf{\hat{x}}-\mathbf{x}_{0}}\right\|_{2}}{\left\|{\mathbf{x}_{0}}\right\|_{2}},
\end{equation}
where $ \mathbf{\hat{x}} $ and $ \mathbf{x}_{0} $ are the denoised signal and the noiseless signal respectively.

We generate a synthetic data set, including 90 random damping complex exponential signals with $2N=256$, $512$, and $1024$ respectively, and repeat 100 Monte Carlo trials to incorporate the randomness of Gaussian noise. Each signal in the data set has $3R+1$ parameters, including $R$, $a_{r}$, $f_{r}$ and $\tau_{r}$, where $r=1,2,\cdots,R$. The number of exponential components is $R=4+M_{r}$, where $M_{r}$ denotes a pseudo-random scalar integer of range $\left[{1,9}\right]$. The amplitude $a_{r}$ is uniformly sampled from $\left({0,10}\right)$. Each frequency $f_{r}$ is uniformly sampled from $\left({0,1}\right)$. The damping factor is ${\tau}_{r}=5+60m_{r}$, where $m_{r}$ is uniformly sampled from $\left({0,1}\right)$.

\begin{figure*}[ht]
  \centering
  \includegraphics[width=5.5in]{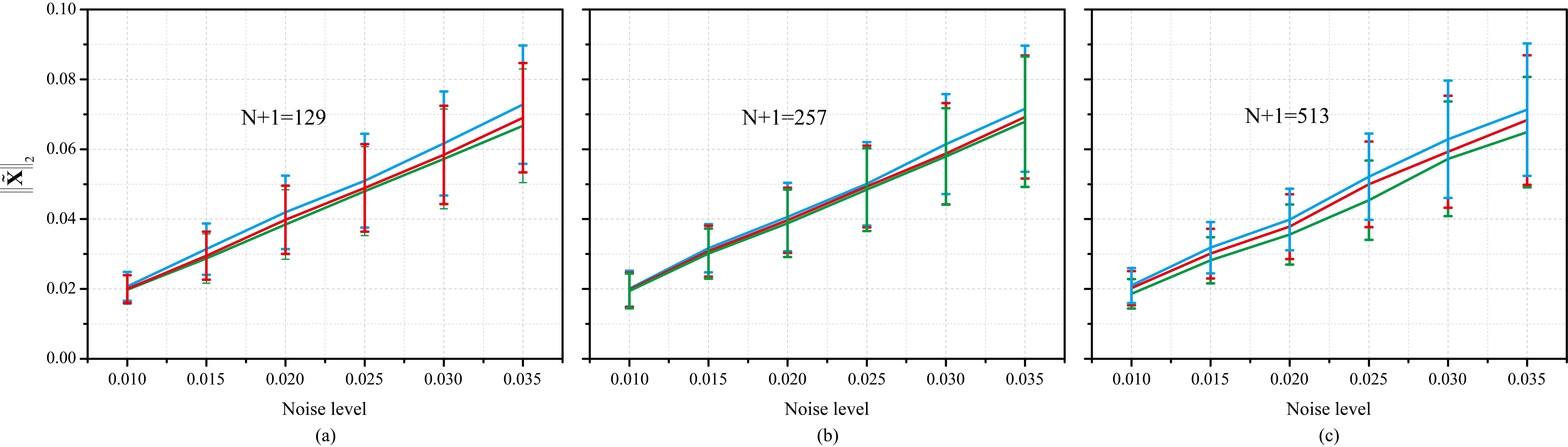}
  \caption{
  $\left\|\mathbf{\tilde{X}}\right\|_2$ with different length, $\sigma$, signals and noises. (a)-(c) show the average $\left\|\mathbf{\tilde{X}}\right\|_2$ of three signals with different length. The matrix $\mathbf{\tilde{X}}$ is of size $N\times\left(N+1\right)$ with $(N+1)=129$, $257$, $513$, respectively. The horizontal axis denotes the standard deviation of the noise, and the vertical axis denotes $\left\|\mathbf{\tilde{X}}\right\|_2$. The vertical error bars represent the randomness of the noises. Note: The setting of signal parameters has been illustrated after the evaluation criteria. 
  }\label{fig:5}
\end{figure*}

Then, we use a series of $\lambda$ to denoise signals in the data set above, find the optimal solution $\mathbf{\hat{x}}$ corresponded to the lowest error, RLNE, and calculate $\left\|\mathbf{\tilde{X}}\right\|_2$. 9 signals with different data lengths are randomly selected and the corresponding $\left\|\mathbf{\tilde{X}}\right\|_2$ are presented in Fig. \ref{fig:5}. 

Results in Fig. \ref{fig:5} indicate that $\mathbb{E}\left\|\mathbf{\tilde{X}}\right\|_2$ is independent of the length and the randomness of the noises and signals. Moreover, this empirical mean of $\left\|\mathbf{\tilde{X}}\right\|_2$ increases as the increase of the standard deviation of the noise. We estimate the slope on Matlab platform and suggest $ \mathbb{E}\left\|\mathbf{\tilde{X}}\right\|_2 = 1.94\sigma $ for denoising.

\section{Numerical Experiments}\label{section:4}

In this section, we evaluate the performance of CHORD with the suggested $\lambda$ on the synthetic data and a realistic NMR spectroscopy data set. 

The typical method, Cadzow \cite{2010_Gillard_CadzowAlgorithm, 2014_PNAS_Delsuc}, and the state-of-the-art method, rQRd \cite{2014_PNAS_Delsuc} are compared with our proposed method. For Cadzow, its key parameter is the rank of this Hankel matrix. For rQRd, its primary parameter is the number of the matrix $\mathbf{Q}$'s column, denoted as $rank_Q$, in QR decomposition. For the rest of the manuscript, without explicit illustration, the main parameters in Cadzow and rQRd are chosen to be the ones yielding the lowest reconstruction error, RLNE.

\subsection{Denoising of synthetic complex data}\label{subsection4.1}

We generated a synthetic exponential complex data with five peaks (presented in Fig. \ref{fig:1}(a)). In the following, the synthetic data indicates the signal in Fig. \ref{fig:1}(a). The denoising performance of three methods is tested through recovering the signal from complex Gaussian noise with different standard deviation ($\sigma=0.01$, $0.02$, $0.03$, $0.04$, $0.05$, $0.06$, $0.07$, and $0.08$, respectively). 100 Monte Carlo trials are done to avoid the randomness of noise.

In practice, we do not know in advance the standard deviation of the noise that corrupts the signal of interest. Here, we use the last 100 time-domain data points of the signal to estimate the standard deviation of the noise to mimic the real cases. Also, we compare the denoised performances of CHORD given the known standard deviation and the estimated standard deviation. For clarity, we name the CHORD using the known standard deviation CHORD$_{\text{Prior}}$ and the CHORD using estimated standard deviation CHORD$_{\text{Esti}}$, respectively.

\begin{figure}[ht]
  \centering
  \includegraphics[width=3.0in]{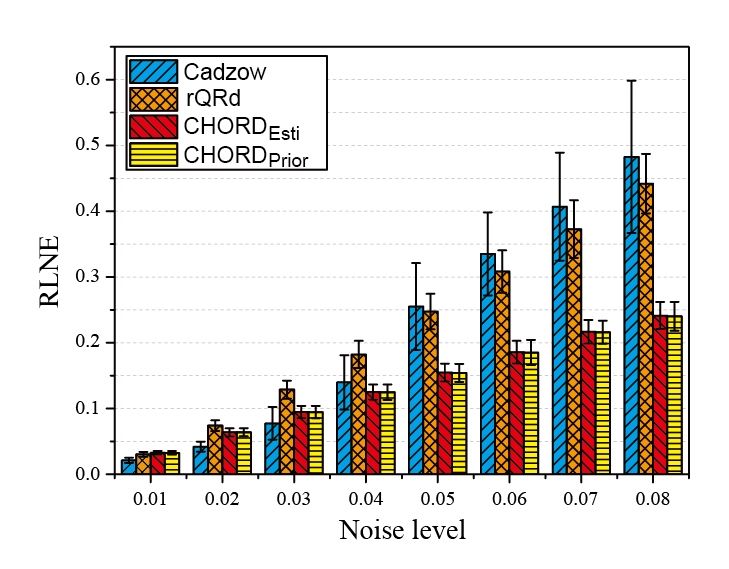}
  \caption{
  The reconstruction error, RLNE, for synthetic data (Fig. \ref{fig:1}(a)) under different noise levels. CHORD$_{\text{Esti}}$ and CHORD$_{\text{Prior}}$ denote denoised results of CHORD with estimated standard deviation and the known standard deviation, respectively. Cadzow and rQRd present the optimal (minimal RLNE) denoised results, respectively. The height of columns shows the average of the RLNEs over 100 trials. The vertical bar comes from the randomeness of noise.
  }\label{fig:6}
\end{figure}

Fig. \ref{fig:6} shows the denoising performance under different noise levels. Under relatively weak noise ($\sigma\leq0.03$), Cadzow achieves the lowest RLNE compared to other approaches. Under relatively high noise ($\sigma\geq0.05$), however, the RLNEs of Cadzow increase faster than that of rQRd and, particularly, CHORD, implying Cadzow is not robust to relatively high noise levels. The proposed method produces the lowest RLNE when the noise is higher than 0.04 and produces smallest variances. Furthermore, the results of CHORD with the estimated noise standard deviation are very close to that of CHORD with known noise standard deviation, indicating the feasibility of CHORD. In the following, without explicit illustration, the mentioned CHORD is CHORD$_{\text{Esti}}$.

\begin{figure}[ht]
  \centering
  \includegraphics[width=3.5in]{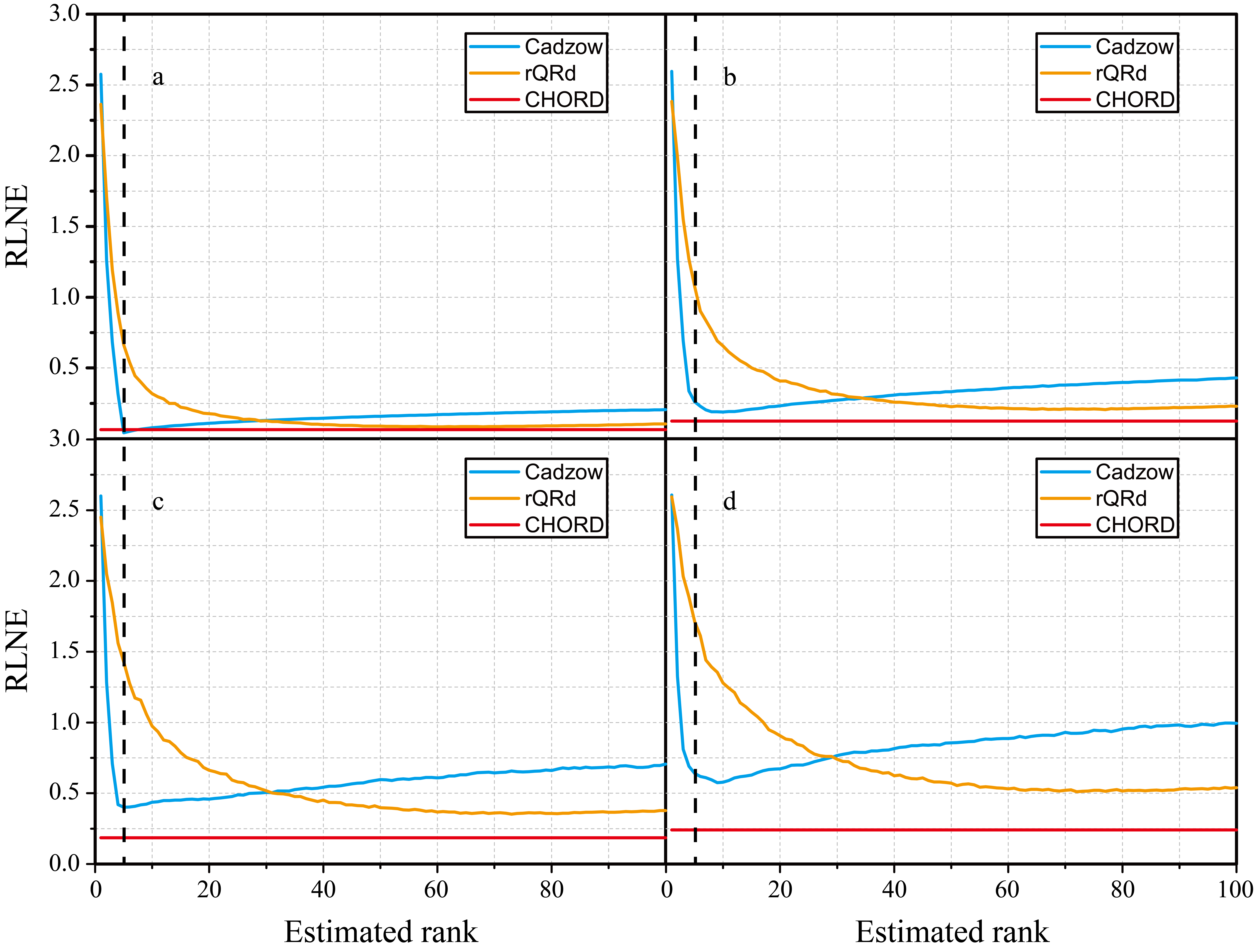}
  \caption{
  The average RLNE of denoised results of the synthetic data (in Fig. \ref{fig:1}(a)) with different estimated ranks over 50 Mont Carlo trials. (a)-(d) denote the average RLNE of denoised results with $\sigma=0.02$, $0.04$, $0.06$, and $0.08$, respectively. The black dash lines stand for the exact rank of the synthetic data (rank=5). Note: For rQRd, the estimated rank stands for $rank_{Q}$.
  }\label{fig:7}
\end{figure}

We evaluate the effect of parameters selection of the tested approaches in Fig. \ref{fig:7}. For Cadzow, when the noise is weak (Fig. \ref{fig:7}(a)), an accurate estimate leads to a good result. But as the noise gets stronger, the optimal estimated rank (in terms of RLNE) may be not equivalent to the actual rank (Fig. \ref{fig:7}(d)), which means that if the noise level is strong enough, an accurate estimated rank will not significantly improve denoised results. Compared with Cadzow, rQRd owns a more flexible parameter setting, but the average RLNE of its denoised results is always higher than that of CHORD under large noise.

\begin{figure*}[ht]
  \centering
  \includegraphics[width=5.3in]{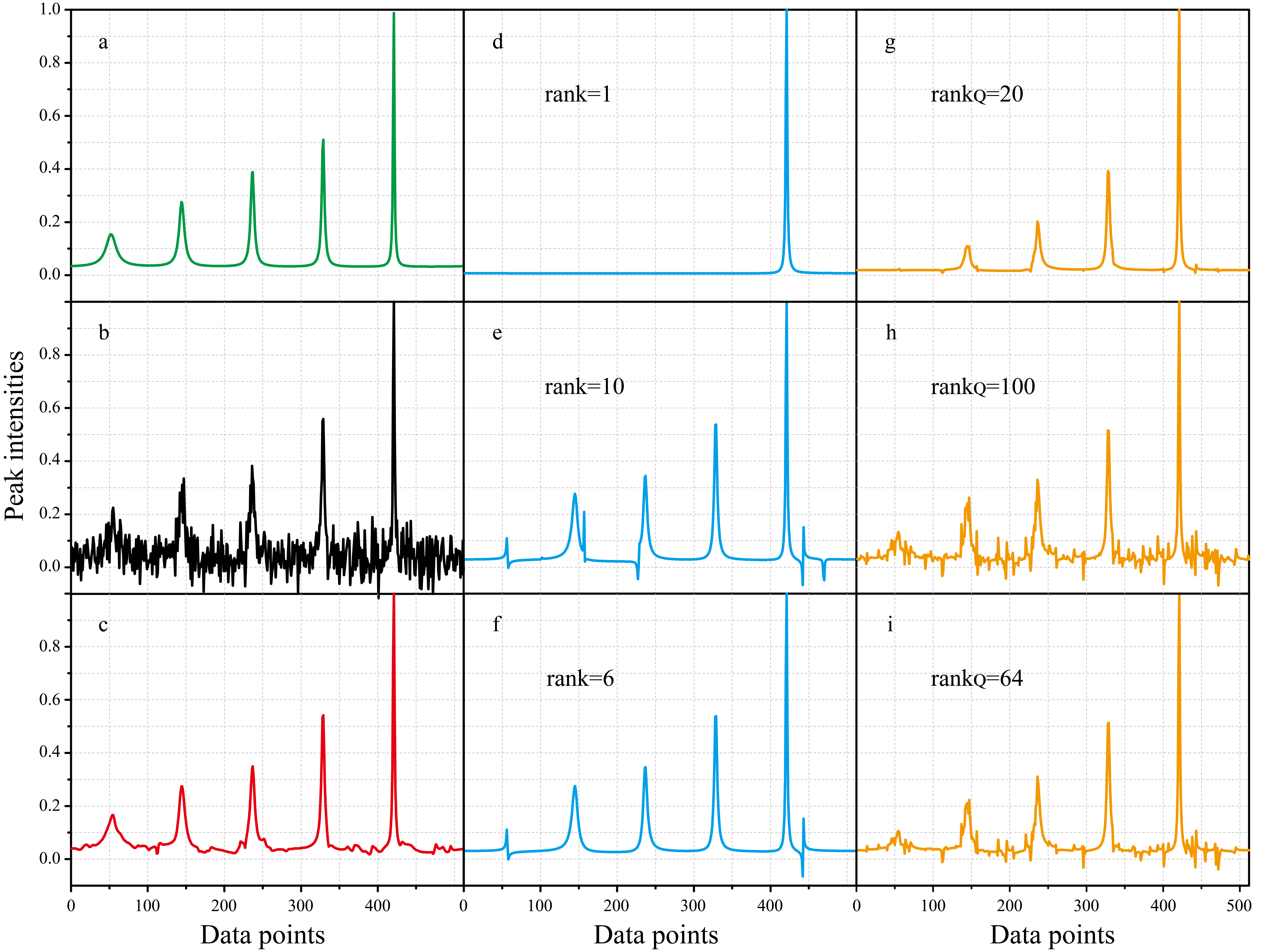}
  \caption{
  The typical denoising results comparison. (a) and (b) denote the synthetic signals without and with noise ($\sigma=0.05$) respectively. (c) is the denoising results of CHORD with the suggested parameter. (d)-(f) show denoised result of Cadzow with three different estimated ranks (small, large, and optimal in terms of RLNE). (g)-(i) are denoised results of rQRd with three different estimated ranks (small, large, and optimal in terms of RLNE). 
  }\label{fig:8}
\end{figure*}

Fig. {\ref{fig:8}} presents the representative denoised results of the synthetic signal corrupted by strong noise. Typical denoised spectra of Cadzow and rQRd with three different parameters selection are presented. Cadzow tends to remove small peaks if using a much smaller estimated rank (see Fig.{\ref{fig:8}}(d)). And if the estimated rank is close to or larger than the real rank, Cadzow spectra introduce spectral distortions and distinct artifacts (see Fig.{\ref{fig:8}}(d) and (f)). For rQRd, a small $rank_{Q}$ leads to a smooth spectrum but with missed or weakened low-intensity peaks (see Fig.{\ref{fig:8}}(g)), while larger parameters introduce strong noise (see Fig.{\ref{fig:8}}(h) and (i)). For the CHORD, it provides a relatively reasonable denoised result using the suggested $\lambda$ and the estimated noise level.

\subsection{Denoising of real NMR spectroscopy data}\label{subsection4.2}

NMR spectroscopy, as a non-invasive technology, has been widely utilized in the study of chemistry, biology, and medicine, such as the diagnosis of diseases \cite{2013_TBME_ZPLiang}. One of the reasons that limits the widespread of this technology is its relatively low SNR. Therefore, CHORD is evaluated on the denoising of a real NMR spectroscopy data. We acquired the signal with high SNR as the reference and add the Gaussian noise retrospectively.

In applications, the unit of chemical shift is usually expressed in part per million (ppm) instead of the Hz, avoiding the ambiguity when spectrometers are at different magnet strengths. The definition of chemical shift is given by

\begin{equation}\label{eq:15}
\text{chemical shift(ppm)}=\frac{f_{test}-f_{ref}}{f_{spec}}\times10^{6},
\end{equation}
where $f_{test}$ denotes the resonance frequency of the sample, $f_{ref}$ the absolute resonance frequency of a standard compound measured in the same magnetic field, and $f_{spec}$ the frequency of the magnetic field strength of spectrometers.

\begin{figure*}[ht]
  \centering
  \includegraphics[width=5.3in]{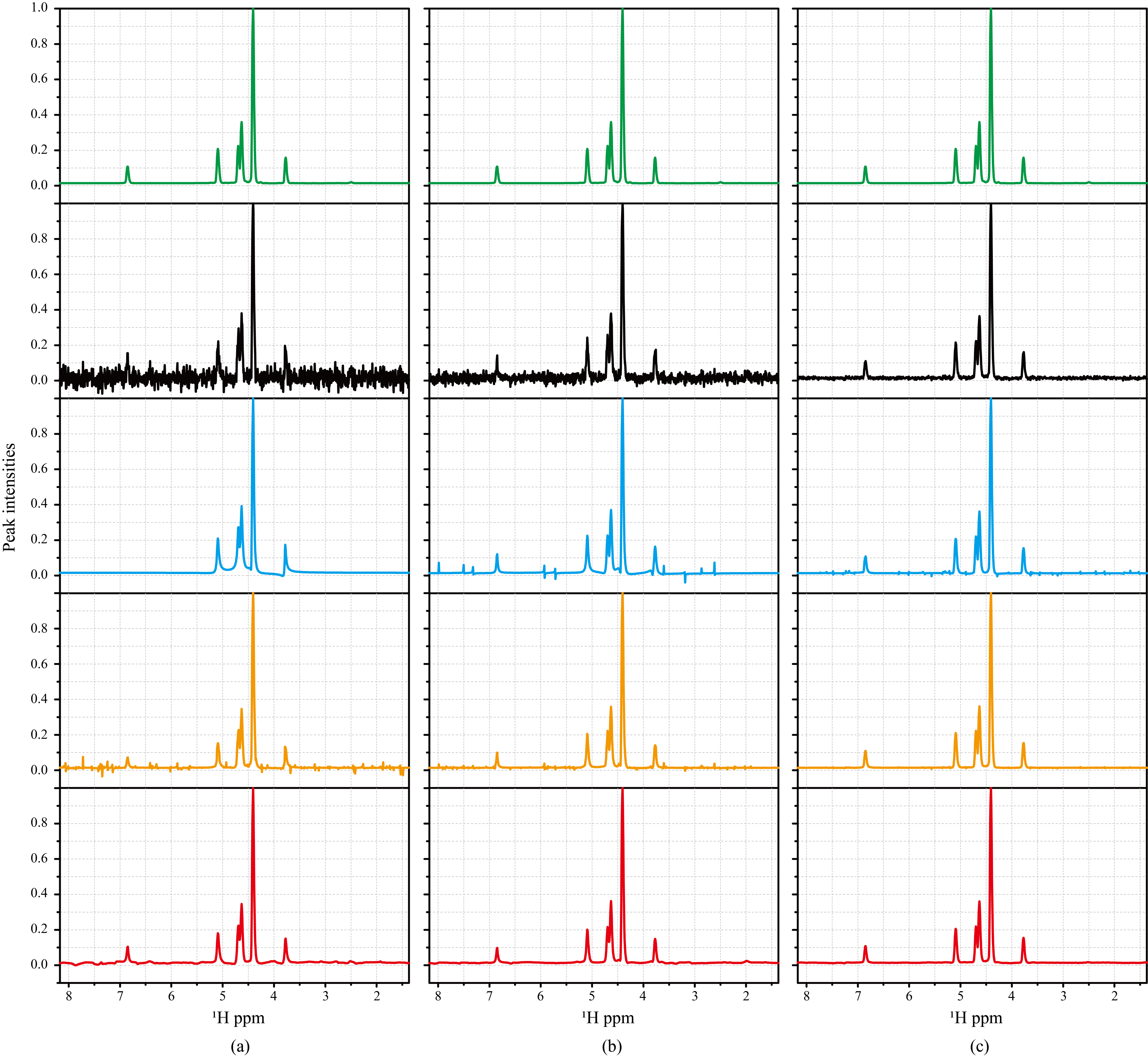}
  \caption{
   Denoised results of a $^1$H spectrum of metabolites with $\sigma=0.035$ (a), $0.020$ (b), and $0.005$ (c), respectively. The green lines denote the ground truth. The black lines indicate observation. The blue, orange and red line are denoised results of Cadzow, rQRd and CHORD, respectively. Note: The results of Cadzow and rQRd that enable the lowest RLNE are presented here.
  }\label{fig:9}
\end{figure*}

The real data is a 1D $^{1}$H NMR spectrum that was acquired at 298 K on a Varian 500 MHz NMR system (Agilent Technologies, Santa Clara,
CA, USA) equipped with a 5 mm indirect detection probe. A standard 1D pulse sequence was used. The experiment time of single scan is 2 s (delay time 1 s and acquisition time 1 s). The sample is a mixture consisting of creatine, choline, magnesium citrate and calcium citrate. The concentration of these metabolites is 2:2:1:1.

The denoised results of the metabolic spectrum are presented in Fig. \ref{fig:9}, which supports the conclusion made on the synthetic data. Under a relatively strong noise level ($\sigma=0.035$), Cadzow smooths the spectrum, which, on the one side, offers a nice noise denoising results, on the other side, however, leads to the missing of some peaks (such as the peaks at 6.8 ppm). rQRd provides a spectrum with obvious noise (orange lines in Fig. \ref{fig:9}(a)), and weakens low-intensity peaks (such as the peaks at 6.8 ppm). CHORD is capable of effectively removing noise and keeping more details of peaks (see Fig. \ref{fig:9}(a)). For the high SNR scenario, all the three methods produce nice and comparable denoised results (see Fig. \ref{fig:9}(c)).

Experiments on synthetic complex exponential and realistic NMR spectroscopy data demonstrate that CHORD with the auto-setting parameter achieves more robust and accurate results compared with Cadzow and rQRd method.

\section{Discussions}\label{section:5}

\subsection{The estimate of noise}\label{subsection5.1}

\begin{figure}[ht]
  \centering
  \includegraphics[width=3.5in]{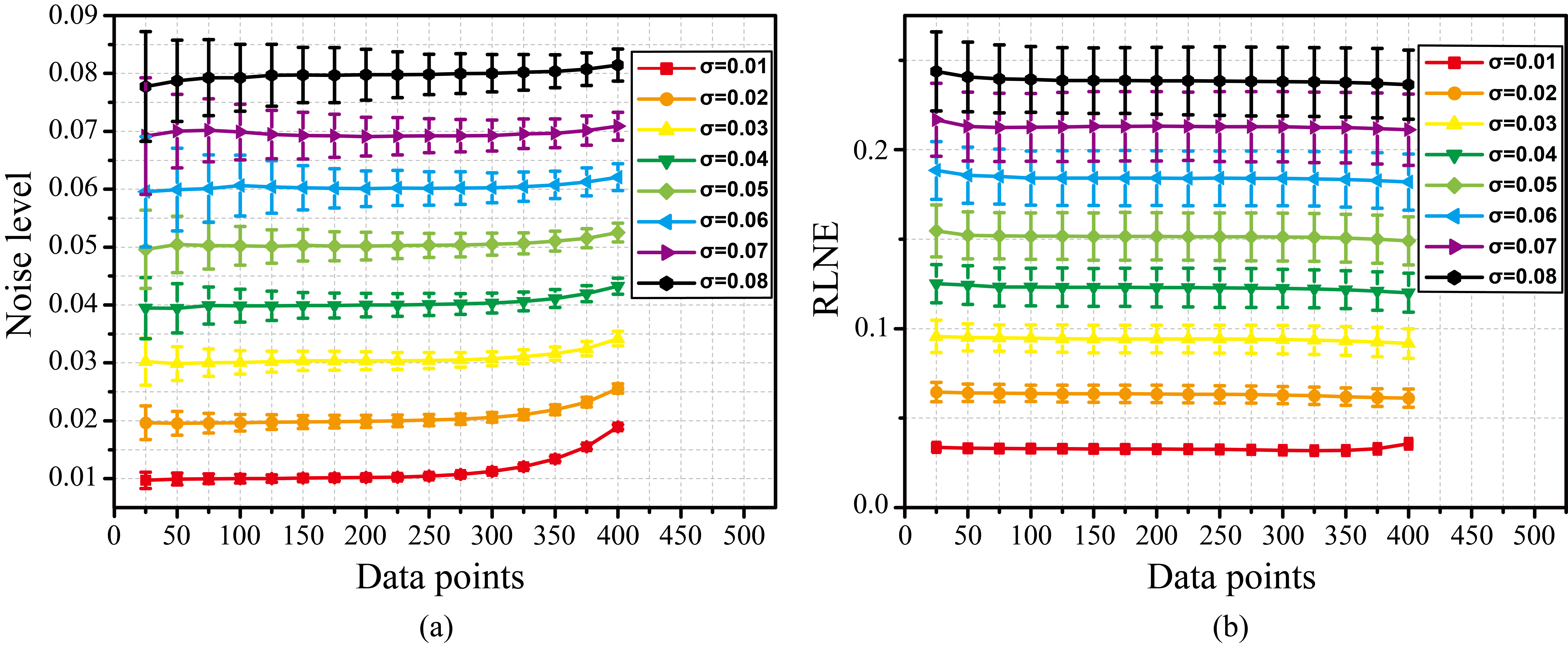}
  \caption{
   The effect of the number of data points. (a) denotes the standard deviation of the estimated noise with different data points. (b) denotes the RLNE of the CHORD denoising results with different estimated noise. The horizontal axis denotes the number of data points utilized in noise estimate. The vertical bars denote the standard deviation of the noise and RLNE, respectively.
  }\label{fig:10}
\end{figure}

We estimate the noise level by calculating the standard deviation of data points at the end of signals on Matlab platform. Ideally, the more data points used to estimate the noise, the better estimation accuracy we can obtain. However, when the noise is relatively large, it is difficult to distinguish signals from noise. Thus, choosing a proper number of data points for noise standard deviation estimation is a challenging task. In this subsection, we discuss the effect of the number of data points used for noise estimate on the denoised results of the synthetic data in Fig. \ref{fig:10}(a). 

We performed experiments with different numbers of data points from the end of the signal to estimate the noise. And then used the estimated standard deviation for spectrum denoising (Fig. {\ref{fig:10}}).

From the results in Fig. {\ref{fig:10} (a)}, we observed that, using fewer data points results in larger vertical bars, while using too many data points causes a larger standard deviation estimate. Besides, for the high SNR signal, the estimate is sensitive to the selection of the number of data points (Fig. {\ref{fig:10}}(a), $\sigma=0.01$, $0.02$). Therefore, we recommend to use the last 100 data points. Notably, the results in Fig. {\ref{fig:10}}(b) indicate that the number of data points used for noise estimation makes no distinction on the denoised results (in terms of RLNE) except in the high SNR case (red lines in Fig. {\ref{fig:10}}(b)).

\section{Conclusion}\label{section:6}

Based on CHORD, a denoising method based on low-rank Hankel property of complex exponential signals, we attempt to figure out the bound of the regularization parameter, determine the empirical optimal constant, and estimate the standard derivation of the noise, so that the users are able to apply CHORD with a auto-setting parameter. Experiments on synthetic complex exponential and realistic NMR spectroscopy data demonstrate that CHORD with the auto-setting parameter achieves more robust and accurate results compared with Cadzow and rQRd method.

In this paper, we did not discuss the effect of $\mathbf{x}_{0}$ and have not provided a theoretical estimate of $\mathbb{E}\left\|\mathbf{\tilde{X}}\right\|_2$. For the future work, it is worthwhile to explore an accurate estimate of $\mathbb{E}\left\|\mathbf{\tilde{X}}\right\|_2$. Furthermore, we are also interested in exploring the probability distribution of the spectral norm, and extending the 1-D model in (\ref{eq:2}) to higher dimensional signals since their acquisition costs relatively more time in applications.


%

\appendices

\section*{Acknowledgment}

The authors would like to thank Hengfa Lu for polishing writing and Zhangren Tu for preparing part of code for comparison.

\ifCLASSOPTIONcaptionsoff
  \newpage
\fi



%



\bibliographystyle{IEEEtran}
\bibliography{./References}

\begin{thebibliography}{1}
\providecommand{\url}[1]{#1}
\csname url@samestyle\endcsname
\providecommand{\newblock}{\relax}
\providecommand{\bibinfo}[2]{#2}
\providecommand{\BIBentrySTDinterwordspacing}{\spaceskip=0pt\relax}
\providecommand{\BIBentryALTinterwordstretchfactor}{4}
\providecommand{\BIBentryALTinterwordspacing}{\spaceskip=\fontdimen2\font plus
\BIBentryALTinterwordstretchfactor\fontdimen3\font minus
  \fontdimen4\font\relax}
\providecommand{\BIBforeignlanguage}[2]{{%
\expandafter\ifx\csname l@#1\endcsname\relax
\typeout{** WARNING: IEEEtran.bst: No hyphenation pattern has been}%
\typeout{** loaded for the language `#1'. Using the pattern for}%
\typeout{** the default language instead.}%
\else
\language=\csname l@#1\endcsname
\fi
#2}}
\providecommand{\BIBdecl}{\relax}
\BIBdecl

\bibitem{2005_LAIA_Tonge}
I.~Masri and A.~Tonge, ``Norm estimates for random multilinear {H}ankel
  forms,'' \emph{Linear Algebra and Its Applications}, vol. 402, no.~1, pp.
  255--262, 2005.

\bibitem{1995_Kavsin_Lowerbound}
B.~Ka{\v{s}}in and L.~Tzafriri, \emph{Lower estimates for the supremum of some
  random processes, {II}}.\hskip 1em plus 0.5em minus 0.4em\relax
  Max-Planck-Institut f{\"u}r Mathematik, 1995.

\bibitem{2015_FnTML_Tropp}
J.~A. Tropp, ``An introduction to matrix concentration inequalities,''
  \emph{Foundations and Trends{\textregistered} in Machine Learning}, vol.~8,
  no. 1-2, pp. 1--230, 2015.

\end{thebibliography}


\begin{thebibliography}{10}
\providecommand{\url}[1]{#1}
\csname url@samestyle\endcsname
\providecommand{\newblock}{\relax}
\providecommand{\bibinfo}[2]{#2}
\providecommand{\BIBentrySTDinterwordspacing}{\spaceskip=0pt\relax}
\providecommand{\BIBentryALTinterwordstretchfactor}{4}
\providecommand{\BIBentryALTinterwordspacing}{\spaceskip=\fontdimen2\font plus
\BIBentryALTinterwordstretchfactor\fontdimen3\font minus
  \fontdimen4\font\relax}
\providecommand{\BIBforeignlanguage}[2]{{%
\expandafter\ifx\csname l@#1\endcsname\relax
\typeout{** WARNING: IEEEtran.bst: No hyphenation pattern has been}%
\typeout{** loaded for the language `#1'. Using the pattern for}%
\typeout{** the default language instead.}%
\else
\language=\csname l@#1\endcsname
\fi
#2}}
\providecommand{\BIBdecl}{\relax}
\BIBdecl

\bibitem{2009_Nature_Inomata}
K.~Inomata, A.~Ohno, H.~Tochio, S.~Isogai, T.~Tenno, I.~Nakase, T.~Takeuchi,
  S.~Futaki, Y.~Ito, H.~Hiroaki, and M.~Shirakawa, ``High-resolution
  multi-dimensional {NMR} spectroscopy of proteins in human cells,''
  \emph{Nature}, vol. 458, no. 7234, pp. 106--109, 2009.

\bibitem{2010_NatureP_Beckonert}
O.~Beckonert, M.~Coen, H.~C. Keun, Y.~Wang, T.~M.~D. Ebbels, E.~Holmes, J.~C.
  Lindon, and J.~K. Nicholson, ``High-resolution magic-angle-spinning nmr
  spectroscopy for metabolic profiling of intact tissues,'' \emph{Nature
  Protocols}, vol.~5, no.~6, pp. 1019--1032, 2010.

\bibitem{1996_NatureM_Preul}
M.~C. Preul, Z.~Caramanos, D.~L. Collins, J.-G. Villemure, R.~Leblanc,
  A.~OLivier, R.~Pokrupa, and D.~L. Arnold, ``Accurate, noninvasive diagnosis
  of human brain tumors by using proton magnetic resonance spectroscopy,''
  \emph{Nature Medicine}, vol.~2, no.~3, pp. 323--325, 1996.

\bibitem{2014_SSNMR_Man}
P.~P. Man, C.~Bonhomme, and F.~Babonneau, ``Denoising {NMR} time-domain signal
  by singular-value decomposition accelerated by graphics processing units,''
  \emph{Solid State Nuclear Magnetic Resonance}, vol.~61, pp. 28--34, 2014.

\bibitem{2019_TMI_Fan}
\BIBentryALTinterwordspacing
F.~Lam, Y.~Li, and X.~Peng, ``Constrained magnetic resonance spectroscopic
  imaging by learning nonlinear low-dimensional models,'' \emph{IEEE
  Transactions on Medical Imaging}, 2019. [Online]. Available:
  \url{https://doi.org/10.1109/TMI.2019.2930586}
\BIBentrySTDinterwordspacing

\bibitem{2014_PNAS_Delsuc}
L.~Chiron, M.~A.~V. Agthoven, B.~Kieffer, C.~Rolando, and M.-A. Delsuc,
  ``Efficient denoising algorithms for large experimental datasets and their
  applications in {F}ourier transform ion cyclotron resonance mass
  spectrometry,'' \emph{Proceedings of the National Academy of Sciences}, vol.
  111, no.~4, pp. 1385--1390, 2014.

\bibitem{2014_MRM_ZPLiang}
F.~Lam and Z.-P. Liang, ``A subspace approach to high-resolution spectroscopic
  imaging,'' \emph{Magnetic Resonance in Medicine}, vol.~71, no.~4, pp.
  1349--1357, 2014.

\bibitem{2018_TSP_Ying}
J.~Ying, J.-F. Cai, D.~Guo, G.~Tang, Z.~Chen, and X.~Qu, ``Vandermonde
  factorization of {H}ankel matrix for complex exponential signal
  recovery—{A}pplication in fast {NMR} spectroscopy,'' \emph{IEEE
  Transactions on Signal Processing}, vol.~66, no.~21, pp. 5520--5533, 2018.

\bibitem{1997_TBME_Tufts}
Y.~Lu, S.~Joshi, and J.~M. Morris, ``Noise reduction for {NMR} {FID} signals
  via {G}abor expansion,'' \emph{IEEE Transactions on Biomedical Engineering},
  vol.~44, no.~6, pp. 512--528, 1997.

\bibitem{1988_TASSP_Cadzow}
J.~A. Cadzow, ``Signal enhancement-{A} composite property mapping algorithm,''
  \emph{IEEE Transactions on Acoustics, Speech, and Signal Processing},
  vol.~36, no.~1, pp. 49--62, 1988.

\bibitem{1993_JMR_LPHwang}
Y.-Y. Lin and L.-P. Hwang, ``{NMR} signal enhancement based on matrix property
  mappings,'' \emph{Journal of Magnetic Resonance, Series A}, vol. 103, no.~1,
  pp. 109--114, 1993.

\bibitem{2010_Gillard_CadzowAlgorithm}
J.~Gillard, ``Cadzow’s basic algorithm, alternating projections and singular
  spectrum analysis,'' \emph{Statistics and Its Interface}, vol.~3, no.~3, pp.
  335--343, 2010.

\bibitem{2015_Angew_Xiaobo}
X.~Qu, M.~Mayzel, J.-F. Cai, Z.~Chen, and V.~Orekhov, ``Accelerated {NMR}
  spectroscopy with low-rank reconstruction,'' \emph{Angewandte Chemie
  International Edition}, vol.~54, no.~3, pp. 852--854, 2015.

\bibitem{2019_Anie_Qu}
\BIBentryALTinterwordspacing
X.~Qu, Y.~Huang, H.~Lu, T.~Qiu, D.~Guo, T.~Agback, V.~Orekhov, and Z.~Chen,
  ``Accelerated nuclear magnetic resonance spectroscopy with deep learning,''
  \emph{Angewandte Chemie International Edition}, 2019. [Online]. Available:
  \url{https://doi.org/10.1002/anie.201908162}
\BIBentrySTDinterwordspacing

\bibitem{2013_TBME_ZPLiang}
H.~M. Nguyen, X.~Peng, M.~N. Do, and Z.~P. Liang, ``Denoising {MR}
  spectroscopic imaging data with low-rank approximations,'' \emph{IEEE
  Transactions on Biomedical Engineering}, vol.~60, no.~1, pp. 78--89, 2013.

\bibitem{2019_MRI_Noiseworthy}
A.~Santos-D{\'i}az and M.~D. Noseworthy, ``Comparison of compressed sensing
  reconstruction algorithms for $^31${P} magnetic resonance spectroscopic
  imaging,'' \emph{Magnetic Resonance Imaging}, vol.~59, pp. 88--96, 2019.

\bibitem{2016_MRM_Cao}
P.~Cao, P.~J. Shin, I.~Park, C.~Najac, I.~Marco‐Rius, D.~B. Vigneron, S.~J.
  Nelson, S.~M. Ronen, and P.~E.~Z. Larson, ``Accelerated high‐bandwidth {MR}
  spectroscopic imaging using compressed sensing,'' \emph{Magnetic Resonance in
  Medicine}, vol.~76, no.~2, pp. 369--379, 2016.

\bibitem{1980_PNMR_Lindon}
J.~C. Lindon and A.~G. Ferrige, ``Digitisation and data processing in {F}ourier
  transform {NMR},'' \emph{Progress in Nuclear Magnetic Resonance
  Spectroscopy}, vol.~14, no.~1, pp. 27--66, 1980.

\bibitem{1995_TIT_Donoho}
D.~L. Donoho, ``De-noising by soft-thresholding,'' \emph{IEEE Transactions on
  Information Theory}, vol.~41, no.~3, pp. 613--627, 1995.

\bibitem{1997_JMR_Barache}
D.~Barache, J.-P. Antoine, and J.-M. Dereppe, ``The continuous wavelet
  transform, an analysis tool for {NMR} spectroscopy,'' \emph{Journal of
  Magnetic Resonance}, vol. 128, no.~1, pp. 1--11, 1997.

\bibitem{1990_PNAS_Hoch}
D.~L. Donoho, I.~M. Johnstone, A.~S. Stern, and J.~C. Hoch, ``Does the maximum
  entropy method improve sensitivity?'' \emph{Proceedings of the National
  Academy of Sciences}, vol.~87, no.~13, pp. 5066--5068, 1990.

\bibitem{2015_ARNMRS_Takeda}
K.~Takeda, ``Solid-state covariance {NMR} spectroscopy,'' \emph{Annual Reports
  on NMR Spectroscopy}, vol.~84, pp. 77--113, 2015.

\bibitem{2007_BBAB_Glaubitz}
C.~Kaiser, J.~J. Lopez, W.~Bermel, and C.~Glaubitz, ``Dual transformation of
  homonuclear solid-state {NMR} spectra—an option to decrease measuring
  time,'' \emph{Biochimica et Biophysica Acta (BBA)-Biomembranes}, vol. 1768,
  no.~12, pp. 3107--3115, 2007.

\bibitem{1977_AC_Malinowski}
E.~R. Malinowski, ``Determination of the number of factors and the experimental
  error in a data matrix,'' \emph{Analytical Chemistry}, vol.~49, no.~4, pp.
  612--617, 1977.

\bibitem{1999_JCACS_Malinowski}
------, ``Abstract factor analysis of data with multiple sources of error and a
  modified {F}aber-{K}owalski f-test,'' \emph{Journal of Chemometrics: A
  Journal of the Chemometrics Society}, vol.~13, no.~2, pp. 69--81, 1999.

\bibitem{2019_ASR_Laurent}
G.~Laurent, W.~Woelffel, V.~Barret-Vivin, E.~Gouillart, and C.~Bonhomme,
  ``Denoising applied to spectroscopies--part {I}: concept and limits,''
  \emph{Applied Spectroscopy Reviews}, vol. 1-29, 2019.

\bibitem{2014_TIT_Chi}
Y.~Chen and Y.~Chi, ``Robust spectral compressed sensing via structured matrix
  completion,'' \emph{IEEE Transactions on Information Theory}, vol.~60,
  no.~10, pp. 6576--6601, 2014.

\bibitem{1996_Hoch_NMR}
J.~C. Hoch and A.~S. Stern, \emph{{NMR} {D}ata {P}rocessing}.\hskip 1em plus
  0.5em minus 0.4em\relax Wiley-Liss New York, 1996.

\bibitem{1999_PNMRS_Koehl}
P.~Koehl, ``Linear prediction spectral analysis of {NMR} data,'' \emph{Progress
  in Nuclear Magnetic Resonance Spectroscopy}, vol.~34, no. 3-4, pp. 257--299,
  2009.

\bibitem{2010_SIAM_JFCai}
J.-F. Cai, E.~J. Cand{\`e}s, and Z.~Shen, ``A singular value thresholding
  algorithm for matrix completion,'' \emph{SIAM Journal on Optimization},
  vol.~20, no.~4, pp. 1956--1982, 2010.

\bibitem{2011_FTM_Boyd}
S.~Boyd, N.~Parikh, E.~Chu, B.~Peleato, and J.~Eckstein, ``Distributed
  optimization and statistical learning via the alternating direction method of
  multipliers,'' \emph{Foundations and Trends{\textregistered} in Machine
  learning}, vol.~3, no.~1, pp. 1--122, 2011.

\bibitem{1992_Watson_LAA}
G.~A. Watson, ``Characterization of the subdifferential of some matrix norms,''
  \emph{Linear Algebra and Its Applications}, vol. 170, pp. 33--45, 1992.

\bibitem{2003_Bertsekas_Conv}
D.~P. Bertsekas, A.~Nedic, and A.~E. Ozdaglar, \emph{Convex {A}nalysis and
  {O}ptimization}.\hskip 1em plus 0.5em minus 0.4em\relax Athena Scientific,
  2003.

\bibitem{2010_ProIEEE_Candes}
E.~J. Cand{\`{e}}s and Y.~Plan, ``Matrix completion with noise,''
  \emph{Proceedings of the IEEE}, vol.~98, no.~6, pp. 925--936, 2010.

\bibitem{2007_ECP_Meckes}
M.~W. Meckes, ``On the spectral norm of a random {T}oeplitz matrix,''
  \emph{Electronic Communications in Probability}, vol.~12, pp. 315--325, 2007.

\bibitem{2013_Math_Meckes}
V.~V. Nekrutkin, ``Remark on the norm of random hankel matrices,''
  \emph{Vestnik St. Petersburg University: Mathematics}, vol.~46, pp. 189--192,
  2013.

\bibitem{2015_FnTML_Tropp}
J.~A. Tropp, ``An introduction to matrix concentration inequalities,''
  \emph{Foundations and Trends{\textregistered} in Machine Learning}, vol.~8,
  no. 1-2, pp. 1--230, 2015.

\bibitem{2015_TSP_Wenjing}
W.~Liao, ``{MUSIC} for multidimensional spectral estimation: stability and
  super-resolution,'' \emph{IEEE Transactions on Signal Processing}, vol.~63,
  no.~23, pp. 6395--6406, 2015.

\end{thebibliography}

%




\end{document}


\title{Supplementary}

\author{Tianyu~Qiu, Wenjing~Liao, Di~Guo, Dongbao~Liu, Xin~Wang, JianFeng~Cai and~Xiaobo~Qu $^{*}$}

\markboth{IEEE TRANSACTIONS ON BIOMEDICAL ENGINEERING,~Vol.~, No.~, ~}%
{Shell \MakeLowercase{\textit{et al.}}: Bare Demo of IEEEtran.cls for IEEE Journals}

\maketitle

\IEEEpeerreviewmaketitle

\section{The proof of Theorem 1}{\label{appendix:A}}

The proof of Theorem 1 are based on the following lemmas:

  \begin{lemma}\label{lemma:1}
  Suppose the real and imaginary parts of the entries in $ \mathbf{z}{\in}\mathbb{C}^{2N+1} $ are i.i.d. Gaussian random variables with mean 0 and variance $ {\sigma}^{2} $. Define $\{d_k\}_{k=0}^{2N}$ as Theorem 1, and then
  \begin{equation}\label{eq:16}
    \left\|{\mathbf{Z}}\right\|_{2}\geq \frac{(N+1)\sigma}{2N+1}\sup_{0{\leq}\omega{\leq}1}\left|{\sum_{k=0}^{2N}d_{k}p_{k}e^{i2{\pi}k\omega}}\right|,
  \end{equation}
  where $ p_{k}\sim \mathcal{N}(0,1), k=0,1,2,\cdots,2N $.
\end{lemma}

\begin{proof}

  For any vectors $ \mathbf{a}, \mathbf{b} \in \mathbb{C}^{N+1}$,
$$
    \left|{\langle{\mathbf{b},\mathbf{Za}}\rangle}\right|=\left|{\mathbf{b}^{H}\mathbf{Za}}\right|
    \leq\left\|{\mathbf{b}}\right\|_{2}\left\|{\mathbf{a}}\right\|_{2}\left\|{\mathbf{Z}}\right\|_{2}.
$$
  We use the technique in \cite{2005_LAIA_Tonge} to derive the lower bound of $\mathbb{E}{\left\|{\mathbf{Z}}\right\|_{2}}$ by choosing proper vectors $\mathbf{a}$ and $\mathbf{b}$. Let $ a_{k_1}=\frac{1}{k_{1}+1}e^{i2{\pi}k_{1}\omega} $ and $ b_{k_2}=\frac{1}{k_{2}+1}e^{-i2{\pi}k_{2}\omega} $, where $\omega\in\left[{0,1}\right]$ and $k_{1},k_{2}=0,\ldots,N$. Then

\begin{footnotesize}
  \begin{equation}\label{eq:17}
  \begin{aligned}
    \left\|{\mathbf{Z}}\right\|_{2}&\geq\frac{1}{C_{N}}\sup_{0\leq{\omega}\leq{1}}
    \left|{\sum_{k_{1}=0}^{N}\sum_{k_{2}=0}^{N}\frac{1}{k_{1}+1}\frac{1}{k_{2}+1}}\right.
    \left.{\frac{e^{i2{\pi}k_{1}\omega}e^{i2{\pi}k_{2}\omega}z_{k_{1}+k_{2}}}{w_{k_{1}+k_{2}}}
    }\right| \\
    &=\frac{1}{C_{N}}\sup_{0\leq{\omega}\leq{1}}\left|{\sum_{k=0}^{2N}d_{k}z_{k}e^{i2{\pi}k\omega}}\right|\\
    &=\frac{\sigma}{C_{N}}\sup_{0\leq{\omega}\leq{1}}\left|{\sum_{k=0}^{2N}d_{k}p_{k}e^{i2{\pi}k\omega}}\right|,
    \end{aligned}
  \end{equation}
  \end{footnotesize}
  where
  $C_N=\sum_{k_{1}=0}^{N}\frac{1}{(k_{1}+1)^{2}}
  \leq{\frac{2N+1}{N+1}}$, which yields the conclusion in (\ref{eq:16}).
\end{proof}

\begin{lemma}\label{lemma:2}
  Let $\{d_{k}\}_{k=0}^{2N}$ and $\{p_{k}\}_{k=0}^{2N}$ be the sequences defined in Theorem 1 and Lemma {\ref{lemma:1}}. If $R_N$ and $Q_N$ are defined as 
  $$R_{N}^{2}=\sum_{k=0}^{2N}\left|{d_k}\right|^{2} \text{and } Q_{N}^{4}=\sum_{k=0}^{2N}\left|{d_k}\right|^{4},$$ 
  then there exists a constant $C$ such that
  \begin{equation}\label{eq:18}
    \mathbb{E}\left({\sup_{0{\leq}\omega{\leq}1}\left|{\sum_{k=0}^{2N}d_{k}p_{k}e^{i2{\pi}k\omega}}\right|}\right)
    \geq{C}\sqrt{R_{N}^{2}\left(1+\log\frac{R_{N}^{4}}{Q_{N}^{4}}\right)},
  \end{equation}
\end{lemma}

Lemma \ref{lemma:2} is a special case of Theorem in \cite{1995_Kavsin_Lowerbound}. Details of the proof is shown as below.

Here is the proof of Lemma \ref{lemma:2}.

\begin{prop}\cite{1995_Kavsin_Lowerbound}
  For every $M<\infty$ there exists a constant $C(M)>0$ such that, whenever $\left\{{\psi}_{k}\right\}_{k=0}^{2N}$ is a system of functions in an $L_2(\mu)$-space satisfying

  ($1^{\circ}$) $\left\|{\psi_{k}}\right\|_{L_2(\mu)}=1$ and $\left\|{\psi_{k}}\right\|_{L_3(\mu)}{\leq}M$, for all $0{\leq}k{\leq}2N$,

  ($2^{\circ}$) $\left\|{\sum_{k=0}^{2N}d_{k}\psi_{k}}\right\|_{L_2(\mu)}{\leq}M\sqrt{\sum_{k=0}^{2N}\left|d_k\right|^2}$, for all $0{\leq}k{\leq}2N$,

  and $\left\{p_k\right\}_{k=0}^{2N}$ are independent random variables over a probability space $\left(T,\mathcal{T},\tau\right)$ with

  ($3^{\circ}$) $\mathbb{E}\left(p_k\right)=0$, $\mathbb{E}\left|p_k\right|^2=1$, and $\sqrt[3]{\mathbb{E}\left|p_k\right|^3}{\leq}M$, for all $0{\leq}k{\leq}2N$,

  then, for any choice of the coefficients of $\left\{d_k\right\}_{k=0}^{2N}$, we have
  $$
   \mathbb{E}\left\|{\sum_{k=0}^{2N}d_{k}p_{k}\psi_{k}}\right\|_{L_{\infty}(\mu)}
  {\geq}C\sqrt{\sum_{k=0}^{2N}\left|{d_k}\right|^2}
  \sqrt{1+\log\frac{\left(\sum_{k=0}^{2N}\left|{d_k}\right|^2\right)^2}{\sum_{k=0}^{2N}\left|{d_k}\right|^4}}.
  $$
\end{prop}

\begin{proof}
  According to (\ref{eq:17}), $\psi_{k}=e^{i2{\pi}k\omega}$. It is obvious that for all $0{\leq}k{\leq}2N$,

  \begin{equation}\label{eq:19}
    \left\|{e^{i2{\pi}k\omega}}\right\|_{L_2(\mu)}=1 {\text{ and }} \left\|{e^{i2{\pi}k\omega}}\right\|_{L_3(\mu)}=1.
  \end{equation}

  According to the triangle inequality,
  \begin{equation}\label{eq:20}
    \left\|{\sum_{k=0}^{2N}d_{k}\psi_{k}}\right\|_{L_2(\mu)}{\geq}\sqrt{\int_{0}^{1}\sum_{k=0}^{2N}\left|{d_{k}e^{i2{\pi}k\omega}}\right|^{2}d\omega}=\sqrt{\sum_{k=0}^{2N}\left|{d_{k}}\right|^{2}}.
  \end{equation}

  $p_k$ is a random variable which satisfies normal distribution, thus

  \begin{equation}\label{eq:21}
    \mathbb{E}\left(p_k\right)=0,\mathbb{E}\left|p_k\right|^2=1 \text{ and }\sqrt[3]{\mathbb{E}\left|p_k\right|^3}=0.
  \end{equation}

  Combining (\ref{eq:19}), (\ref{eq:20}) and (\ref{eq:21}) yields Lemma \ref{lemma:2}.

\end{proof}

Combining Lemma {\ref{lemma:1}} and Lemma {\ref{lemma:2}} results in Theorem 1.

\section{The proof of Theorem 2}{\label{appendix:B}}

\begin{proof}

We express $\mathbf{Z}$ as a sum of independent matrices such that
  \begin{equation}\label{eq:22}
  \mathbf{Z}=\sum_{k=0}^{2N}\frac{1}{w_{k}}z_{k}\mathbf{B}_{k},
\end{equation}
where $\mathbf{B}_{k}\in\mathbb{R}^{(N+1)\times(N+1)}$ has one on the $(k+1)^{th}$ skew diagonal and all other entries are $0$. For example,
$$ \mathbf{B}_{1}=\left[{
\begin{array}{cccc}
  0 & 1 & \cdots & 0 \\
  1 & 0 & \cdots & 0 \\
  \vdots & \vdots & \vdots & \vdots \\
  0 & 0 & \cdots & 0
\end{array}
}\right] .$$
According to \cite{2015_FnTML_Tropp},
\begin{equation}\label{eq:23}
  \mathbb{E}\left\|{\mathbf{Z}}\right\|_{2}{\leq}\sqrt{2{\nu}^{2}\left({\mathbf{Z}}\right)\log\left({2N+2}\right)},
\end{equation}
where ${\nu}^{2}\left({\mathbf{Z}}\right)=\max\left\{{\left\|{\mathbb{E}\left(\mathbf{Z}^{H}\mathbf{Z}\right)}\right\|_{2},\left\|{\mathbb{E}\left(\mathbf{Z}\mathbf{Z}^{H}\right)}\right\|_{2}}\right\}$. The parameter ${\nu}^{2}\left({\mathbf{Z}}\right)$ can be calculated as follows
\begin{equation}\label{eq:24}
  \begin{aligned}
    \left\|{\mathbb{E}\left(\mathbf{Z}^{H}\mathbf{Z}\right)}\right\|_{2}
    &=\left\|{\mathbb{E}\left(\mathbf{Z}\mathbf{Z}^{H}\right)}\right\|_{2}\\
    &=\left\|{\mathbb{E}\left(
    \left({\sum_{k=0}^{2N}\frac{z_{k}}{w_{k}}\mathbf{B}_{k}}\right)\left({\sum_{m=0}^{2N}\frac{z_{m}}{w_{m}}\mathbf{B}_{m}}\right)^{H}
    \right)}\right\|_{2}\\
    &=\left\|{\mathbb{E}\left(
    \sum_{k=0}^{2N}\frac{\left|{z_{k}}\right|^{2}}{w_{k}^{2}}\mathbf{B}_{k}\mathbf{B}_{k}^{H}
    \right)}\right\|_{2}.
  \end{aligned}
\end{equation}

Denote the diagonal matrix $ \mathbf{C}_{k}=\mathbf{B}_{k}\mathbf{B}_{k}^{H} \in\mathbb{R}^{(N+1)\times(N+1)}$, $k=0,1,\cdots,2N$. When $0{\leq}k{\leq}N$, the first $k+1$ diagonal entries of $\mathbf{C}_{k}$ are one, and others are zero. When $N+1{\leq}k{\leq}2N$, the last $2N+1-k$ diagonal entries of $\mathbf{C}_{k}$ are one, and others are zero. For example,
$$ \mathbf{C}_{1}=\left[
\begin{array}{cccc}
  1 &  &  &  \\
   & 1 &  &  \\
   &  & \ddots &  \\
   &  &  & 0
\end{array}
\right] \text{ and }
 \mathbf{C}_{N+1}=\left[
\begin{array}{cccc}
  0 &  &  &  \\
   & 1 &  &  \\
   &  & \ddots &  \\
   &  &  & 1
\end{array}
\right].$$

Substituting (\ref{eq:24}) into the definition of ${\nu}^{2}\left({\mathbf{Z}}\right)$ in (\ref{eq:23}) results in
\begin{equation}\label{eq:25}
    {\nu}^{2}\left({\mathbf{Z}}\right)={\left\|{\mathbb{E}\left(\sum_{k=0}^{2N}\frac{\left|{z_{k}}\right|^{2}}{w_{k}^{2}}\mathbf{C}_{k}\right)}\right\|_{2}}
    ={\sigma}^2C_{\mathbf{w}}.
\end{equation}
Finally we combine (\ref{eq:25}) and (\ref{eq:23}) to obtain Theorem 2.

\end{proof}

\section{Asymptotic analysis of the estimates of bounds}{\label{appendix:C}}

\begin{theorem}\label{theorem:3}
Let $R_N$ and $Q_N$ be defined as Theorem 1. Then there exists a constant $C_{L}>0$ such that
\begin{equation}\label{eq:26}
  \lim_{N\to{+\infty}}\frac{(N+1){C}}{2N+1}\sqrt{R_{N}^{2}\left({1+\log\frac{R_{N}^{4}}{Q_{N}^{4}}}\right)}=C_{L}.
\end{equation}
\end{theorem}

The proof of Theorem \ref{theorem:3} is based on two lemmas below which study the asymptotic of $R_N^{2}$ and ${Q_{N}^{4}}$ as $N \rightarrow \infty$.

\begin{lemma}\label{lemma:3}
  Let $R_{N}$ be defined as $R_{N}^{2}=\sum_{k=0}^{2N}\left|{d_k}\right|^{2} $. Then there exists a constant $C_{R}>0$ such that
  \begin{equation}\label{eq:27}
    \lim_{N\to{+\infty}}R_{N}^{2}=C_{R}.
  \end{equation}
\end{lemma}

\begin{lemma}\label{lemma:4}
  Let $Q_{N}$ be defined as $Q_{N}^{4}=\sum_{k=0}^{2N}\left|{d_k}\right|^{4}$. Then there exists a constant $C_{Q}>0$ such that
  \begin{equation}\label{eq:28}
    \lim_{N\to{+\infty}}Q_{N}^{4}=C_{Q}.
  \end{equation}
\end{lemma}

Lemma {\ref{lemma:3}} and Lemma {\ref{lemma:4}} are proved as following.

Here is the proof of Lemma {\ref{lemma:3}}.

\begin{proof}
According to the definition, $R_{N}^{2}$ is expressed as

\begin{equation}\label{eq:29}
  \begin{aligned}
    R_{N}^{2}&=4\underbrace{\sum_{k=0}^{N}\left({\frac{1}{(k+1)(k+2)}\sum_{m=0}^{k}\frac{1}{m+1}}\right)^2}_{R_{N}^{(1)}}\\
    &+4\underbrace{\sum_{k=N+1}^{2N}\left({\frac{1}{(2N-k+1)(k+2)}\sum_{m=k}^{2N}\frac{1}{m-N+1}}\right)^2}_{R_{N}^{(2)}}.
  \end{aligned}
\end{equation}

The sequence $R_{N}^{(1)}$ is positive, and increases as $N$ increases. It is straight forward

\begin{equation}\label{eq:30}
  1\leq\sum_{m=0}^{k}\frac{1}{m+1}{\leq}k+1.
\end{equation}

Let $F_{1}(N)=\int_{0}^{N}\frac{1}{\left({l+1}\right)^{2}\left({l+2}\right)^{2}}dl$. The sequence $R_{N}^{(1)}$ satisfies the following upper bound and lower bound

\begin{equation}\label{eq:31}
  \begin{aligned}
    R_{N}^{(1)}&{\geq}\sum_{k=0}^{N}\frac{1}{\left({k+1}\right)^{2}\left({k+2}\right)^{2}}>F_{1}(N)\\
    &=\frac{3}{2}-2\log{2}-\left({\frac{1}{N+1}+\frac{1}{N+2}}\right)\\
    &+2\log\left({1+\frac{1}{N+1}}\right),
  \end{aligned}
\end{equation}
and
\begin{equation}\label{eq:32}
  R_{N}^{(1)}{\leq}\sum_{k=0}^{N}\frac{1}{\left({k+2}\right)^{2}}<1-\frac{1}{N+2}.
\end{equation}

Combining (\ref{eq:31}) and (\ref{eq:32}) gives rise to

\begin{equation}\label{eq:33}
  \frac{3}{2}-2\log{2}<\lim_{N\to{+\infty}}R_{N}^{(1)}<1.
\end{equation}

Since $R_{N}^{(1)}$ increases as $N$ increases,

\begin{equation}\label{eq:34}
  \lim_{N\to{+\infty}}R_{N}^{(1)}=C_{R_1},
\end{equation}
where $\frac{3}{2}-2\log2<C_{R_1}<1$.

The sequence $R_{N}^{(2)}$ can be rewritten as

  \begin{equation}\label{eq:35}
    R_{N}^{(2)}=\sum_{k=0}^{N-1}\left({\frac{1}{(N-k)(k+N+3)}\sum_{m=0}^{N-1-k}\frac{1}{m+k+2}}\right)^2.
  \end{equation}

For $\left({\sum_{m=0}^{N-1-k}\frac{1}{m+k+2}}\right)^{2}$ where $k=0,1,\cdots,N-1$, it is obvious that

$$\left({\sum_{m=0}^{N-1-k}\frac{1}{m+k+2}}\right)^{2}{\leq}\left({\sum_{m=0}^{N-1}\frac{1}{m+2}}\right)^{2}.$$

According to Cauchy-Buniakowsky-Schwarz inequality,
\begin{equation}\label{eq:36}
\begin{aligned}
  \left({\sum_{m=0}^{N-1}\frac{1}{m+2}}\right)^{2}&{\leq}\sum_{m=0}^{N-1}\left({\frac{1}{m+2}}\right)^2\sum_{m=0}^{N-1}1\\
  &<N\left({1-\frac{1}{N+1}}\right)\\
  &<N.
\end{aligned}
\end{equation}

We have

\begin{equation}\label{eq:37}
  0<\sum_{m=0}^{N-1-k}\frac{1}{m+k+2}<\sqrt{N}.
\end{equation}

Denote $g(k)=\frac{N}{\left({N-k}\right)^{2}\left({k+N+3}\right)^{2}}$ where $k=0,1,2,\cdots,N-1$.

The limit of $\sum_{k=0}^{N-1}g(k)$ can be calculated as

\begin{equation}\label{eq:38}
\begin{aligned}
  &\sum_{k=0}^{N-1}\frac{N}{\left({N-k}\right)^{2}\left({k+N+3}\right)^{2}}\\
  &=\sum_{k=0}^{N-1}\frac{2N}{(2N+3)^{3}(N-k)}+\frac{2N}{(2N+3)^{3}(k+N+3)}\\
  &+\sum_{k=0}^{N-1}\frac{N}{(2N+3)^{2}(N-k)^{2}}+\frac{N}{(2N+3)^{2}(k+N+3)^{2}}\\
  &<\frac{4N^{2}}{(2N+3)^{3}}+\frac{2N-1}{(2N+3)^{3}}+\frac{N^2}{(N+1)(N+2)(2N+3)^{2}}
\end{aligned},
\end{equation}
thus
\begin{equation}\label{eq:39}
    \lim_{N\to{+\infty}}\sum_{k=0}^{N-1}g(k)=0.
\end{equation}

Since $R_{N}^{(2)}\geq0$ and it satisfies $R_{N}^{(2)}<\sum_{k=0}^{N-1}g(k)$,

\begin{equation}\label{eq:40}
  \lim_{N\to{+\infty}}R_{N}^{(2)}=0.
\end{equation}

Combining (\ref{eq:33}) and (\ref{eq:39}) results in Lemma \ref{lemma:3}.
\end{proof}

Lemma {\ref{lemma:4}} is proved is below.

\begin{proof}
According to the definition, $Q_{N}^{4}$ is given by

\begin{equation}\label{eq:41}
  \begin{aligned}
    Q_{N}^{4}&=16\underbrace{\sum_{k=0}^{N}\left({\frac{1}{(k+1)(k+2)}\sum_{m=0}^{k}\frac{1}{m+1}}\right)^4}_{Q_{N}^{(1)}}\\
    &+16\underbrace{\sum_{k=N+1}^{2N}\left({\frac{1}{(2N-k+1)(k+2)}\sum_{m=k}^{2N}\frac{1}{m-N+1}}\right)^4}_{Q_{N}^{(2)}}.
  \end{aligned}
\end{equation}

The same technique as Lemma \ref{lemma:3} to prove (\ref{eq:41})

\begin{equation}\label{eq:42}
  \frac{111}{8}-20\log{2}\leq\lim_{N\to{+\infty}}Q_{N}^{(1)}<\frac{1}{3},
\end{equation}
and the details will not be shown here.

The sequence $Q_{N}^{(2)}$ is restated as

  \begin{equation}\label{eq:43}
    Q_{N}^{(2)}=\sum_{k=0}^{N-1}\left({\frac{1}{(N-k)(k+N+3)}\sum_{m=0}^{N-1-k}\frac{1}{m+k+2}}\right)^4.
  \end{equation}

According to (\ref{eq:35}),
\begin{equation}\label{eq:44}
\begin{aligned}
  \left({\sum_{m=0}^{N-1-k}\frac{1}{m+k+2}}\right)^{4}&{\leq}\left({\sum_{m=0}^{N-1}\frac{1}{m+2}}\right)^{4}\\
  &<N^2\\
  &<(k+N+3)^3.
\end{aligned}
\end{equation}

\begin{equation}\label{eq:45}
  0<\left(\sum_{m=0}^{N-1-k}\frac{1}{m+k+2}\right)^{4}<(k+N+3)^3.
\end{equation}

Define $h(k)=\frac{1}{\left({N-k}\right)^{4}\left({k+N+3}\right)}$ where $k=0,1,2,\cdots,N-1$, then the limit of $\sum_{k=0}^{N-1}\frac{1}{\left({N-k}\right)^{4}\left({k+N+3}\right)}$ can be calculated as
\begin{equation}\label{eq:46}
\begin{aligned}
  &\sum_{k=0}^{N-1}\frac{1}{\left({N-k}\right)^{4}\left({k+N+3}\right)}\\
  &=\sum_{k=0}^{N-1}\frac{1}{(2N+3)^{4}(N-k)}+\sum_{k=0}^{N-1}\frac{1}{(2N+3)^{3}(N-k)^{2}}\\
  &+\sum_{k=0}^{N-1}\frac{1}{(2N+3)^{2}(N-k)^{3}}+\sum_{k=0}^{N-1}\frac{1}{(2N+3)(N-k)^{4}}\\
  &+\sum_{k=0}^{N-1}\frac{1}{(2N+3)^{4}(k+N+3)}\\
  &<\frac{2N}{\left({2N+3}\right)^{4}}+\frac{2N-1}{N(2N+3)^{3}}+\frac{2N-1}{N(2N+3)^{2}}+\frac{2N-1}{N(2N+3)}.
\end{aligned}
\end{equation}

Thus

\begin{equation}\label{eq:47}
  \lim_{N\to{+\infty}}\sum_{k=0}^{N-1}h(k)=0.
\end{equation}

Since sequence $Q_{N}^{(2)}{\geq}0$ and it satisfies $Q_{N}^{(2)}<\sum_{k=0}^{N-1}h(k)$,

\begin{equation}\label{eq:48}
  \lim_{N\to{+\infty}}Q_{N}^{(2)}=0.
\end{equation}

Combining (\ref{eq:42}) and (\ref{eq:48}) yields Lemma \ref{lemma:4}.
\end{proof}

Combining Lemma \ref{lemma:3} and Lemma \ref{lemma:4} gives rise to

\begin{equation}\label{eq:49}
\begin{aligned}
  &\lim_{N\to{+\infty}}\frac{(N+1){C}}{2N+1}\sqrt{R_{N}^{2}\left({1+\log\frac{R_{N}^{4}}{Q_{N}^{4}}}\right)}\\
  &={C}\sqrt{C_{R}\left({1+\log\frac{C_{R}^{2}}{C_{Q}}}\right)}\lim_{N\to{+\infty}}\frac{(N+1)}{2N+1}\\
  &=\frac{C}{2}\sqrt{C_{R}\left({1+\log\frac{C_{R}^{2}}{C_{Q}}}\right)},
\end{aligned}
\end{equation}
which gives rise to Theorem \ref{theorem:3} with $$C_{L}=\frac{C}{2}\sqrt{C_{R}\left({1+\log\frac{C_{R}^{2}}{C_{Q}}}\right)}.$$

When $N$ is large enough, the upper bound and the lower bound only differ by a factor of $\sqrt{\log N}$. According to the analysis above, we suggest to choose $\mathbb{E}\left\|\mathbf{Z}\right\|_{2}$ as

\begin{equation}\label{eq:50}
  \mathbb{E}\left\|\mathbf{Z}\right\|_{2}=\frac{C(N+1)}{(2N+1)}\sqrt{R_{N}^{2}\left({1+\log\frac{R_{N}^{4}}{Q_{N}^{4}}}\right)}{\sigma}.
\end{equation}

\ifCLASSOPTIONcaptionsoff
  \newpage
\fi

\bibliographystyle{IEEEtran}
\bibliography{./References}
